\documentclass[reprint, amsmath,amssymb, aps, nofootinbib, superscriptaddress]{revtex4}

\usepackage{graphicx}  
\usepackage{dcolumn}   
\usepackage{bm}        
\usepackage{amssymb, amsmath}
\usepackage{subcaption}
\usepackage{color,soul}
\usepackage{float}
\usepackage{upgreek}
\usepackage{multirow}
\usepackage[nolist,nohyperlinks]{acronym}
\usepackage{relsize}
\usepackage{xfrac}
\usepackage{mathtools}

\usepackage[latin1]{inputenc}
\usepackage[francais]{babel}
\definecolor{blue}{rgb}{0.36, 0.54, 0.66}
\definecolor{amaranth}{rgb}{0.9, 0.17, 0.31}
\definecolor{pink}{rgb}{0.57, 0.36, 0.51}
\definecolor{ao}{rgb}{0.0, 0.5, 0.0}
\definecolor{maroon}{rgb}{0.76, 0.13, 0.28}
\definecolor{cardinal}{rgb}{0.77, 0.12, 0.23}
\definecolor{frenchlila}{rgb}{0.53, 0.38, 0.56}

\usepackage[pagebackref=false,
			hyperindex=true,
			citecolor=cardinal,
			colorlinks=true,
			linkcolor=blue,
			urlcolor=blue]{hyperref}

\renewcommand{\refeq}[1]{eq.~(\ref{#1})}

\newcommand{\lb}{\left(}
\newcommand{\rb}{\right)}

\newcommand{\bi}{\begin{itemize}}
\newcommand{\ei}{\end{itemize}}
\newcommand{\bfig}{\begin{figure}}
\newcommand{\efig}{\end{figure}}
\newcommand{\ba}{\begin{eqnarray}}
\newcommand{\ea}{\end{eqnarray}}
\newcommand{\be}{\begin{equation}}
\newcommand{\ee}{\end{equation}}
\newcommand{\bea}{\begin{eqnarray}}
\newcommand{\eea}{\end{eqnarray}}
\newcommand{\beq}{\begin{equation}}
\newcommand{\eeq}{\end{equation}}
\newcommand{\beqar}{\begin{eqnarray}}
\newcommand{\eeqar}{\end{eqnarray}}
\newcommand{\beqars}{\begin{eqnarray*}}
\newcommand{\eeqars}{\end{eqnarray*}}
\newcommand{\bc}{\begin{center}}
\newcommand{\ec}{\end{center}}

\newcommand{\tobs}{$\mathrm{T}_\mathrm{obs}$ }

\begin{document}

\preprint{APS/123-QED}


\title{Characterization of the stochastic signal originating from compact binaries populations as measured by LISA}

 \author{Nikolaos  Karnesis}
\affiliation{APC, AstroParticule et Cosmologie, CNRS, Universit\'e de Paris,  F-75013 Paris, France}
\affiliation{Department of Physics, Aristotle University of Thessaloniki, Thessaloniki 54124, Greece}
 \author{Stanislav Babak}
\affiliation{APC, AstroParticule et Cosmologie, CNRS, Universit\'e de Paris,  F-75013 Paris, France}
\affiliation{Moscow Institute of Physics and Technology, Dolgoprudny, Moscow region, Russia}
\author{Mauro Pieroni}
\affiliation{Blackett Laboratory, Imperial College London, SW7 2AZ, UK}
\author{Neil Cornish}
\affiliation{eXtreme Gravity Institute, Department of Physics, Montana State University, Bozeman, Montana 59717, USA}
\author{Tyson Littenberg}
\affiliation{NASA Marshall Space Flight Center, Huntsville, AL 35812, USA}

\date{\today}

%
%
%
%
%
%
\begin{abstract}
The Laser Interferometer Space Antenna (LISA) mission,  scheduled for launch in the early 2030s,  is a gravitational wave observatory in space designed to detect sources emitting in 
the milli-Hertz band. In contrast to the present ground based detectors, the LISA data are expected to be a signal-dominated, with strong and 
weak gravitational wave signals overlapping in time and in frequency. Astrophysical population models predict a sufficient number of signals in the LISA band 
to blend together and form an irresolvable foreground noise. In this work, we present a generic method for characterizing the 
foreground signals originating from a given astrophysical population of coalescing compact binaries. Assuming idealized detector conditions 
and perfect data analysis technique capable of identifying and removing the bright sources, we apply an iterative procedure which allows 
us to predict the different levels of foreground noise.
\end{abstract}

\maketitle

%
%
%
%
%
%
\section{Introduction
\label{sec:intro}}

The Laser Interferometer Space Antenna (LISA) is a space-based Gravitational Wave (GW)
observatory mission scheduled for launch in the early 2030s~\cite{2017arXiv170200786A}. 
Unlike currently observed GW signals, LISA sources are expected to be long-lived, with thousands of strong GW signals present in the data overlapping in time and/or in frequency. 
This implies that all GW signals from various sources will be have to be fit for and characterized simultaneously. The sources generating overlapping 
signals include Super-Massive Black Hole Binaries 
(SMBHBs)~\cite{PhysRevLett.96.111102, berti2006}, Stellar-mass Black Hole Binaries 
(SBBHs)~\cite{ligocatalogue, PhysRevLett.118.121101, PhysRevLett.122.081102, PhysRevD.91.022003, sesanasobbhs, Abbott2020mjq}, 
ultra-Compact Binaries originating in our Galaxy (CGBs)~\cite{Hiscock_2000, Kalogera_2001, Nelemans2001hp, Marsh_2011}, 
Extreme Mass Ratio Inspirals (EMRIs)~\cite{PhysRevD.94.124042, PhysRevD.86.104050, Chua_2015, Barack2005aj}, and 
a Stochastic GW Background (SGWB) that may originate from cosmological sources~\cite{Caprini_2018, PhysRevD.79.062002}. 
The number and density of GW sources in the LISA band presents a data analysis challenge of producing a \emph{global} fit~\cite{Robson_2017}, simultaneously detecting and classifying overlapping signals, which is often referred to as the ``source confusion problem.''   

Astrophysical population models predict a sufficiently large number of sources to form a resolution-limited GW foreground signal that could be measured by LISA even after
all resolvable sources have been fit and removed from the data. The unresolved component is usually referred to as ``confusion noise'', following initial studies that employed a resolvability criterion based on source density, demanding that the number of data points equal or exceed the number of parameters required to fit each source~\cite{1997CQGra..14.1439B,2004PhRvD..70l2002B}. Later studies showed that a reduction in signal-to-noise due to the ``self-noise'' from other sources in the population generally has a greater impact on detectability~\cite{Timpano:2005gm}, but the terms ``confusion noise'' and ``foreground confusion'' had already taken root, and we will continue to use them here.

Given the expected LISA sensitivity~\cite{scird},  we expect the GW confusion signal from millions of CGBs
to dominate over instrumental noise between ${\sim} 0.4$ and ${\sim} 3$~$\mathrm{mHz}$. These CGBs mostly comprise
detached and interacting white dwarfs~\cite{Stroeer_2006, Marsh_2011}. A small number of them (${\sim} 30$)
have been already observed electromagnetically, and parameters of those binaries suggest that they should emit GWs in the LISA band. 
A subset of them should be easily detectable, and those binaries are referred to as {\em verification binaries}: guaranteed LISA sources 
which can be used to monitor performance of the instruments~\cite{10.1093/mnras/sty1545}. Models for the full population of CGBs suggests upwards of ${\sim} 60$ million binaries \cite{Nelemans_2004, Korol_2018, Korol_2017} in the LISA band, with only a small 
fraction (${\sim} 10$ thousand) of them being individually resolvable. Thus, most of those binaries will contribute to the confusion foreground signal.
 
Another guaranteed population of binaries in the local Universe consists of SBBHs.
Merging SBBH are observed by the LIGO and Virgo detectors, with over 40 such systems 
having been reported after partial analysis of data from the 3rd observational run~\cite{ligoo3}. Those binaries at the early stage of their inspiral emit a GW signal at the high frequency end of LISA measurement band.  While it was suggested that it might be 
challenging to recover such sources with LISA~\cite{mooregerosaklein}, the ensemble of these sources could potentially 
be another source of a confusion foreground for LISA. So far, upper bounds on the stochastic signals have been estimated by a direct 
extrapolation from the LIGO-Virgo event rate~\cite{Abbott_2007,PhysRevD.93.042005,PhysRevD.91.022003,
PhysRevLett.118.121101, PhysRevLett.122.081102}.

At very low frequencies we expect a stochastic GW signal from a population of SMBHBs in sufficiently wide orbits to be emitting almost monochromatic GW signals.  The SMBHB background is the one of the primary sources targeted  
Pulsar Timing Arrays, which use radio emission from the utra-stable millisecond pulsars to detect GWs in the 
nano-Hz band. 
The upper limits on the GW amplitude  are reported in Refs.~\cite{Arzoumanian_2016, Arzoumanian_2018, 
PhysRevD.92.102003, Lentati2015qwp}, and  recent results \cite{Nanograv12.5} suggest the presence 
of a common red-noise process among the observed pulsars which might be confirmed as a GW signal in the near future. 
The merger of SMBHBs will be observed by LISA, and detection of stochastic GW signal by PTA might 
constrain the expected event rate of those sources in the LISA band. 



The presence of astrophysical foregrounds poses a data analysis challenge for the LISA mission. Naturally, the combination of the different 
foreground and background signals will complicate the source separation procedures for LISA. 
In addition, the exact level of the confusion noise will affect the achievable precision of source parameter estimation for resolvable signals. 
Thus, it is of great importance to study, understand, and predict the overall shape and 
amplitudes of the potential foreground components of the LISA data. This will allow us to make more realistic predictions 
for the resolvable sources with LISA, as well as to gain a better understanding on the data analysis requirements for
the mission.  On the other hand, the separation and characterization of the different contributions to the foreground signal may provide additional insight into the channels of formation and evolution of astrophysical populations that generate those signals.
For example, current astrophysical uncertainties for some sources translate to several orders of magnitude uncertainty in the event rate, to wit, predicted event rates for 
EMRIs are between a few to a few thousands per year \cite{BabakEMRIs}, making the presence or absence of an EMRI-induced foreground an informative constraint on the underlying theory. 

In this work we present a generic method for predicting and characterization the foreground signals as measured by LISA. We begin with a combination of ideas from previous works ~\cite{Timpano:2005gm,Nissanke_2012,Cornish2017vip,PhysRevD.75.043008}, upon which we expand in two directions. 
First, we use updated estimates of the  instrument capabilities, based on the state-of-the-art studies of the LISA design~\cite{scird}, 
and we simulate and analyze the data directly on the LISA TDI~\cite{tdi, aet} data streams. We have used both simple 
analytic models for the instrument noise power spectral density (PSD), and time-series simulated using {\tt LISA Code}~\cite{lisacode}. Secondly, 
as described in section~\ref{sec:methodology}, we have developed a framework to perform our analysis for different 
types of foregrounds, generated by different types of binary sources. 
Then, in sections~\ref{sec:gbs} and~\ref{sec:sobbhs}, we apply this methodology, as an illustrative example, 
to the foreground confusion signal generated by the CGBs and SBBHs respectively. 
As a further demonstration, given a model to characterize the residual foreground signal, we perform a Markov Chain Monte Carlo (MCMC) 
analysis to infer parameters of the foreground signal which encodes information about the underlying population. 
We also demonstrate that resolving the bright CGB and SBBH sources leads to the separability of their respective foreground signals despite the GWs being emitted by the underlying populations overlapping in frequency between ${\sim} 2$ to ${\sim} 5$~$\mathrm{mHz}$. 
Finally, in section~\ref{sec:comb} we study how the presence of two astrophysical foreground components affect the parameter estimation of 
resolvable signals using a Fisher Information Matrix approach.

\begin{figure*}[tbh]
\centering
\begin{subfigure}{.5\textwidth}
  \centering
  \includegraphics[width=1.\linewidth]{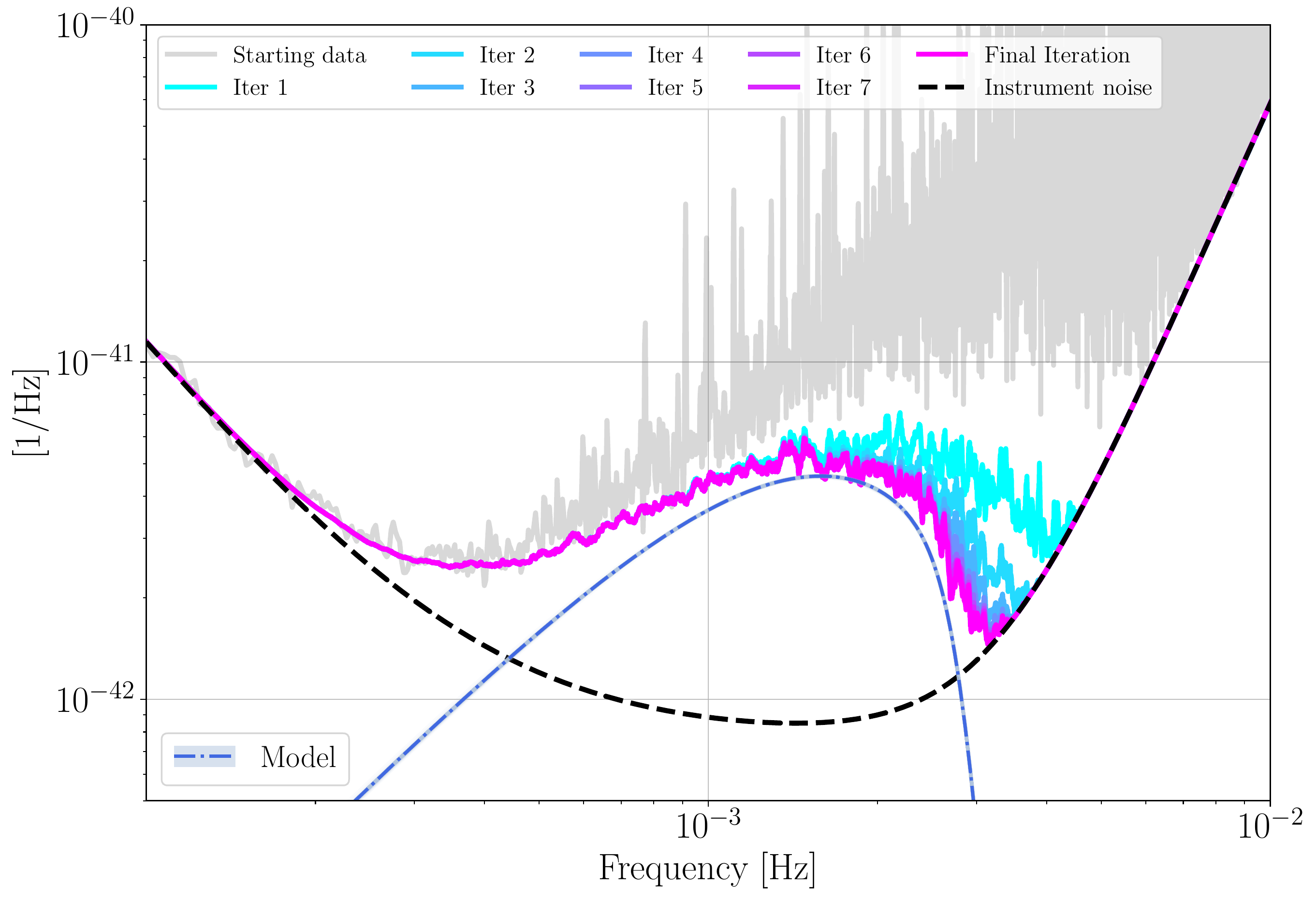}
  \caption{}
  \label{fig:gbsalgo}
\end{subfigure}%
\begin{subfigure}{.5\textwidth}
  \centering
  \vspace{-4mm} 
  \includegraphics[width=.98\linewidth]{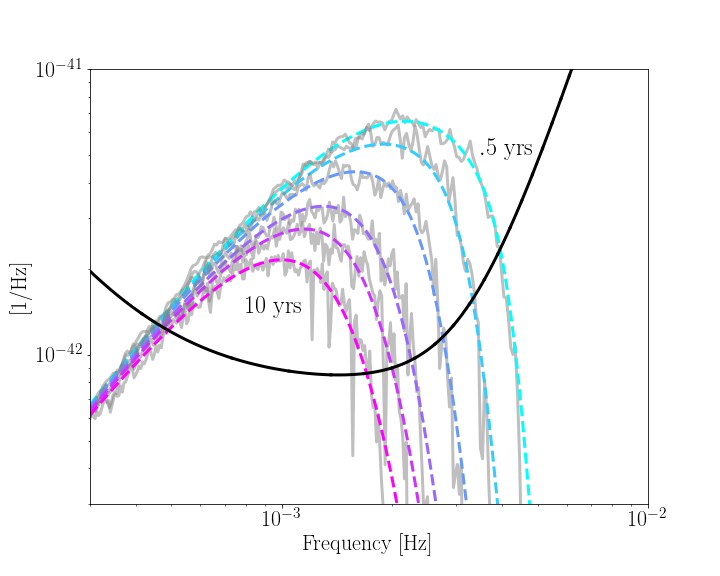}
  \caption{}
  \label{fig:galfit}
\end{subfigure}
\caption{(a): Illustration of the progress of the procedure for estimating the confusion noise. 
		For the population of the CGBs described in section \ref{sec:gbs}, it takes 6 to 10 iterations 
		for the algorithm to converge, depending on the convergence tolerance criterion. The starting data are represented with the gray curve 
		(300 segment averaged spectrum), while the final combined instrument and confusion noise is represented here in bright magenta. The dark blue dashed line represents 
		the analytic model fit of the confusion noise component, together with the associated error-bars. (b): Using the analytical model of 
		\refeq{eq:galfittobs}, and the parameter values from table~\ref{tab:galmodel}, we can make a prediction of the level of the confusion noise 
		due to CGBs, depending on a given observation time. Here, we depict the power spectrum of the residual data with blue, while the 
		orange curves represent the model prediction for the given \tobs.}
\label{fig:galbinariesall}
\end{figure*}

\section{Methodology\label{sec:methodology}}

The methodology we followed is based on the rather strong assumption that bright sources with a Signal-to-Noise Ratio (SNR) larger than 
a given threshold are detected and characterized without systematic bias or source confusion (see Ref.~\cite{Robson_2017} for more refined estimates that take into account imperfect signal removal). The detection threshold is adjusted 
for the different source types, taking into account search and parameter estimation studies, 
as for example in~\cite{tysonpipe} for CGBs and~\cite{mooregerosaklein} for SBBHs. 
Assuming ideal noise conditions, where the spectral properties of the readout and test-mass are identical in each link, we define the total SNR estimate for a given source as
\be
\rho_\mathrm{tot}^2 = \sum_K  \lb h_K | h_K \rb,
\label{eq:snrtot}
\ee  
with $K \in \{A, E \}$ the noise-orthogonal TDI variables~\cite{tdi,aet}, 
and $\lb \cdot | \cdot \rb$ denotes the noise weighted inner product expressed for two time 
series $a$, and $b$ as
\be
\lb a | b \rb = 2 \int\limits_0^\infty \mathrm{d}f \left[ \tilde{a}^\ast(f) \tilde{b}(f) + \tilde{a}(f) \tilde{b}^\ast(f) \right]/\tilde{S}_n(f),
\label{eq:ineerprod} 
\ee
where the tilde represents the data in Fourier frequency domain, and the asterisk stands for the complex conjugate.
The $\tilde{S}_n(f)$ is one-sided PSD of the noise for the different TDI variables $S_K(f)$.

One can estimate the residual confusion noise from a population of GW sources by 
subtracting the sources whose SNR exceeds a given detection threshold. The problem arises when one has
to compute~\refeq{eq:snrtot}, where $\tilde{S}_n(f)\equiv\tilde{S}_\mathrm{instr}(f) + \tilde{S}_\mathrm{conf}(f)$ 
includes the confusion residual GW signal $\tilde{S}_\mathrm{conf}(f)$, and the instrumental noise $\tilde{S}_\mathrm{instr}(f)$. A solution is to follow 
a methodology similar to~\cite{Timpano:2005gm}, and set-up an iterative process to estimate $\tilde{S}_n(f)$.
The basic steps of the analysis are summarized as
\bi
	\item[a.] First, the data generation takes place either in frequency or in time domain, based on a predefined observation time 
	and a given sampling frequency. We simulate the GW signals from the population of $N$ sources drawn from a given astrophysical population model.
		During this process we also compute the SNR of each source with respect to the adopted instrumental noise model, we refer 
		to it as  the optimal Signal-to-Noise Ratio (SNR) of the source in isolation, $\rho^\mathrm{iso}_i$.  We will use $\rho^\mathrm{iso}_i$ as a measure of GW strength in the next step.
	\item[b.] We estimate the confusion noise $S_{n,\,k}$ using either a running mean or median 
		on the power spectrum of the data. The index $k$  refers to the iteration number. On top of the smoothed PSD, 
		one can also fit a polynomial model, or perform spline interpolation smoothing (which we usually do). Then, assuming the  SNR threshold $\rho_0$, 
		we calculate the SNR, $\rho_i$, for each source  $i \in N$ using the smoothed $S_{n,\,k}$ as the PSD in \refeq{eq:ineerprod}.
        If $\rho_i > \rho_0$, the source is subtracted from the data. In this step we make use of the previously calculated $\rho^\mathrm{iso}_i$ to accelerate the procedure:	
		if $\rho^\mathrm{opt}_i<\varkappa \rho_0$, we skip computation of the SNR for this ($i$-th) source, instead automatically adding it to the confusion. 
		Here $\varkappa \le 1$ is some safety factor. Note, that we assume perfect source identification and subtraction, which will not be achievable in practice~\cite{Robson_2017}.
		
	\item[c.] After  subtracting the brightest sources in the previous step, we return back to
		the step (a) where we evaluate the smoothed PSD of the residuals  $S_{n,\,k+1}$ and iterate steps (a)-(b).
		The algorithm stops when either no sources exceed the threshold $\rho_0$, or $S_{n,\,k+1}\simeq S_{n,\,k}$ based on a fractional tolerance 
		limit which we choose to be around ${\sim} 5$\% for all frequencies and all channels. In principle, both criteria can be met 
		at the same time. We usually choose to work with the fractional tolerance because it accelerates convergence, since the number of
		sources subtracted at the final iterations of the algorithm is small.
	\item[d.] The process converges within 5-10 iterations resulting in the smooth estimation of the PSD for the confusion noise, $S_\mathrm{final}$. 
	We use it to evaluate the final SNR for all subtracted source during the iterative procedure. In addition, we evaluate a projected accuracy in measuring 
	parameters of those sources based on the Fisher Information matrix (FIM). This iterative scheme is illustrated in figure~\ref{fig:gbsalgo} 
	for the population of CGBs (see next section for details).
	\item[e.] Finally we perform an MCMC analysis to fit a parameterized model for the confusion noise $S_\mathrm{final}$ to the residual.
\ei		 
We find that the choice of the window length for the noise estimation, or the method for the final step of smoothing the residual (using either polynomial or spline fits) does not have significant impact on the result. 
On the other hand, we find noticeable differences between using a running mean or median for the 
case of the PSD smoothing method. The PSD quantity is $\chi^2$-distributed (i.e. has a fatter tail towards higher values), so 
in principle we should expect that even in the absence of signals the smoothing method following a running mean to be larger 
than the running median case. We nevertheless perform analysis using both methods and report below our findings. 

To conclude this section, we comment on the working assumption employed in the definition of the procedure. We have assumed idealized 
conditions for the instrumental noise: the noise is stationarity, and there are no gaps, glitches, or instrumental ``lines" (monochromatic or almost monochromatic signal 
of instrumental or environmental origin). All these artifacts either could be identified and removed \cite{ligoglitch,Robson:2018jly,Cornish:2020dwh,Chatziioannou:2021ezd} and/or will affect the 
SNR threshold for a reliable detection.  Note that we do not need to assume 
that the instrumental noise is known as we estimate it at each step of iteration, and the confusion GW signal is just yet another noise component 
(from the SNR computation point of view). We also could deal directly with Michelson $X, Y, Z$ TDI data streams, evaluate the cross-spectra and 
either diagonalize the noise matrix (producing new $A, E, T$ TDI) or work directly with non-diagonal noise matrix. 
 
 Of greater consequence is our assumption of idealized data analysis that can exactly identify the GW signals above a certain threshold. There is an on-going effort within 
 the LISA Data Challenge project to demonstrate and evaluate the performance of various data analysis techniques. Current results look very 
 encouraging (\cite{tysonpipe, MarsatBakerMBHB, NeilMBHB, 
 ToubianaSBBH}) and we are confident that we will be able to reliably detect GW sources in the LISA data, but nevertheless the procedure here represents an unattainable ideal. 
Our analysis neglects the effect of the instrument and confusion noise on the source identification itself: the best recovered parameters from the model will be different from the true values due to presence of  the noise~\cite{Robson_2017}. Furthermore, the observed catalog will unavoidably have some contamination due to several sources being fit with a single model waveform, especially early on during the LISA observing campaign~\cite{tysonpipe}.
As a result removing the estimated GW sources will inevitably remove a small 
portion of the noise with it and/or leave behind some residual signal~\cite{Robson_2017}. An idea to make our simulation more realistic while still computationally tractable would be to 
build a small Markov chain forming posteriors for the source (or collection of overlapped sources). Another solution would be to estimate 
the covariance matrix of the parameters based on a FIM calculation, and then draw the given ``solution'' for each source from a multivariate 
Gaussian distribution. We relegate these more elaborate approaches to future work and consider the results presented here 
as a zero-order approximation which is fast to run and gives a good starting point for a more detailed analysis.

Last but not least, we want to comment on the simulated data. We have to choose the GW model to simulate the LISA data. Naturally one 
would be tempted to use a simplified and fast wave-form model to compute the GW signal, however the final result might strongly 
depend on the simulated LISA data. For example, use of higher order modes in the simulated signal could significantly affect the conclusion, even if those 
modes are not strong for each individual source. The same concerns the simulation of numerous 
sub-threshold events: in simulating the data we need to include also very weak sources, in other words the catalogue should 
be \emph{complete} and extend to include sources which merge beyond the simulated span of the data. 


%
%
%
%
%
%
\section{Application to the case of Compact Galactic Binaries
\label{sec:gbs}}

In this section we apply our methodology to the population of the CGBs.
We study the properties of the confusion signal, after subtracting the ``loud'' counterparts, as 
a function of observation time \tobs. We assume that the orbital period evolves slowly
and thus can use a very simple model for the GW signal where the phase is decomposed in a Taylor series. The signal in the source frame can be presented as  
\ba
	h_+^\mathrm{s} =& \mathcal{A} \lb 1 + \cos^2 \iota \rb \cos \lb \Phi(t) \rb \\
	h_\times^\mathrm{s} =& 2 \mathcal{A} \cos \iota \sin \lb \Phi(t) \rb ,
\ea
with 
\be
	\Phi(t) = \phi_0 + 2\pi f_0 t + \pi \dot{f}_0 t .
\ee
Here $\phi_0$ is the initial GW phase, and we neglect all but the first frequency derivative 
$\dot{f}_0$ as it is sufficient for evaluation of the foreground signal. Higher-order derivatives are seldom detectable for the assumed CGB population~\cite{Littenberg_2019}.
$\mathcal{A}$ is the GW strain amplitude and $\iota$ the inclination. The signal describes early inspiral 
evolution, however, due to mass transfer in the interacting binaries, the system could also ``outspiral'' (negative frequency derivative). 
We have assumed that the binaries are in the circular orbits, which is expected for vast majority of systems.  
We have also neglected the presence of exo-planets or (in general) a third perturbing body 
\cite{Robson:2018svj,lisaexoplanets}. In principle, it is straightforward to include those systems in our study, but again it does not affect our main 
objective: evaluating the number of resolvable systems and characterizing the residual confusion foreground. 

Next, we need to transform the waveform to the Solar System Barycenter (SSB), by assuming 
the latitude $\beta$ and longitude $\lambda$ ecliptic coordinates of the source, and introducing the polarization 
angle $\psi$. We follow exactly the polarization and angle conventions of~\cite{ldcdoc}, and an implementation based on~\cite{gbwf}. 

We chose to work with the catalogues of the {\em Radler} LISA Data Challenge data-set~\cite{ldcdata, ldcdoc}. 
This catalogue contains the signal of 29857650 CGBs, generated by the population models of \cite{Korol2020l, Nelemans2013yg}.

For a given set of \tobs and SNR threshold $\rho_0$, we generate idealized data-sets directly in the frequency domain. 
The data are simplified in the sense that they do not contain any other type of GW signals.  In addition to the signals we simulated instrumental noise using the LISA Code simulator~\cite{lisacode}.
The PSD of the noise could either be empirically determined from the simulated data or we assume the analytic fit~\cite{ldcdoc}. 
We have ran the iterations described in section~\ref{sec:methodology}
for $\rho_0 = \{5,\, 7\}$, and \tobs$=\{0.5, \,1, \, 2, \, 4, \, 6,\, \, 10\}$ years. At the end of each run we the have set of the 
subtracted sources and smoothed CGB foreground signal.

\paragraph{Recovered compact galactic binaries sources}

As expected, the number of recovered sources greatly depends on the chosen SNR threshold $\rho_0$, and 
the LISA observation time \tobs. For the given LISA configuration~\cite{scird}, and setting $\rho_0 = \{5,\, 7\}$, we 
recover the number of sources displayed in table~\ref{tab:sourcesrecovered}. 
The numbers are reported for two methods of evaluating the smoothed PSD \: running median and running mean. 
\begin{table}[h]
\caption{\label{tab:sourcesrecovered}
The recovered number of sources as a function of the LISA observation time \tobs for two methods of smoothening PSD.}
\bc
\begin{ruledtabular}
\begin{tabular}{c|cc|cc}
 \multirow{3}{*}{\tobs $[\mathrm{years}]$} & \multicolumn{4}{c}{\# of sources recovered} \\ \cline{2-5}
 & \multicolumn{2}{c}{Running mean} & \multicolumn{2}{c}{Running median} \\ \cline{2-5}
 & $\rho_0 = 5$ & $\rho_0 = 7$ & $\rho_0 = 5$ & $\rho_0 = 7$ \\ 
\hline\hline
 $0.5$	 & 8031		& 	4491  & 8850	& 	5111 \\
 $1$	 & 14170	&	8487  & 14918	&	9305 \\
 $2$	 & 22681 	&	14206 & 23849 	&	15441 \\
 $4$	 & 35891	&	22951 & 37684	&	25025 \\
 $6$	 & 45925	&	30097 & 48179	&	32702 \\
\end{tabular}
\end{ruledtabular}
\ec
\end{table}

\paragraph{Confusion noise due to compact galactic binaries} 
We use an empirical analytical model for the estimated confusion foreground given as 
 \be
 S_\mathrm{gal} = \frac{A}{2}  f^{-7/3} e^{-(f/f_1)^\alpha} \lb 1 + \mathrm{tanh}\lb \lb f_\mathrm{knee} - f \rb/f_2 \rb \rb .
 \label{eq:galfit}
 \ee
 The first two terms $\propto Af^{-7/3}$ are the expected spectral shape generated from a population binaries evolving due to GW 
 emission~\cite{Cornish2017vip}, the second factor accounts for the loss of stochasticity as we move to higher frequencies and 
 the factor in the brackets (with hyperbolic tangent) reflects that after (roughly)  $f\ge f_{\rm{knee}}$ we should be able to 
 resolve and subtract all GBs. The fit for each observation time \tobs is shown in figure~\ref{fig:galfit}. 
 
 The six fitting parameters for the confusion noise model are: two scaling frequencies $f_1, f_2$, the knee frequency 
 $f_{\rm{knee}}$,  the amplitude $A$ and a ``smoothness'' parameter $\alpha$. 
 The overall amplitude $A$ is a property of the population of GBs. The frequency $f_2$ and $\alpha$ are expected to be a property 
 of a population \emph{and} the SNR threshold we use. The two remaining parameters $f_1, f_{\rm{knee}}$ are expected to also have 
 a strong dependence on the observation time: indeed the longer we observe, the more CGBs become resolvable and, therefore, could be subtracted.     

We have simulated LISA TDI time series data for a total observation duration of 
$\mathrm{T}_\mathrm{obs,\, max} = 10$ years, and then estimated the confusion noise for different fractions of $\mathrm{T}_\mathrm{obs,\, max}$. 
We have found that the $f_1$ and $f_\mathrm{knee}$ parameters of~\refeq{eq:galfit} can be well approximated as the following functions  
of observation duration \tobs:
\be
\begin{array}{r@{}l}
\mathrm{log}_{10}\lb f_1 \rb &= a_1 \mathrm{log}_{10} (T_\mathrm{obs}) + b_1, \,\, \text{and,} \\
\mathrm{log}_{10}\lb f_\mathrm{knee} \rb &= a_k \mathrm{log}_{10}(T_\mathrm{obs}) + b_k,
\end{array}
\label{eq:galfittobs}
\ee
with $a_1$, $a_k$, $b_1$, and $b_k$ being amplitude calibration parameters. Their values are reported on table~\ref{tab:galmodel}. 
This empirical analytic fit works rather well for observation durations above 1 year and improves with 
the observational duration. The minimum observation time we have checked was 3 months. 
Note that the GW signals from the population of CGBs are cyclo-statationary, meaning that the signal is modulated during the annual orbital motion of LISA: with its antenna pattern 
function pointing toward or away from the Galactic centre where the majority of the sources are found. This non-stationarity is reflected in deviations from the analytic fit 
which depends on where we terminate the observations during the first year. Nonetheless, as seen in fig.~\ref{fig:galfit}, the fit 
is still quite satisfactory.

\paragraph{Fit of the Galactic model to the confusion noise}
As a proof of principle we have also looked at the inverse problem. We have assumed that the numerically evaluated CGB foreground is 
the result of LISA observations (reduced by removing the bright sources). Next, we evaluate the residual signal and infer parameters 
of the simple Galactic model given by Eq.~\eqref{eq:galfit} using Bayesian sampling techniques. For this purpose, we perform a pre-processing 
procedure similar to the one employed in~\cite{Caprini2019pxz, Pieroni2020rob, Flauger2020qyi}. 

Given the full data stream, we first convert the data to $\Omega$ units using:
\begin{equation}
 	h^2 \, \Omega_{GW} \equiv \frac{4 \pi^2 f^3}{3 (H_0 /h)^2 } S(f) \; ,
\end{equation}
$H_0 $ being the Hubble parameter as observed today and $h$ its dimensionless normalization. 
We then cut it into $N_c$ data segments having the same duration of around $11.5$ days corresponding to $10^{-6}$Hz and average over 
segments~\footnote{For a the data set in frequency domain with frequency resolution ${\sim} 1/T_\mathrm{obs}$, for a given frequency 
say $f_i$ this correspond to averaging over data points with $f_i-5\times 10^{-7}$Hz $\leq f \leq f_i+5\times 10^{-7}$Hz.}. At this 
point, in order to further lower the computational complexity of the problem, we coarse-grain (\emph{i.e.} bin) the data to a less dense set (in the 
following denoted $f^{k}_{ij}, D^{k}_{ij}$) by performing an inverse variance weighting of the data using the noise as an estimate 
for the variance (Sec.3.1 of~\cite{Flauger2020qyi} for details). In particular for each decade in frequency above $10^{-4}$Hz we pass 
from the initial $10^{-6}$Hz linear spacing to 250 evenly logarithmically spaced frequency points. Similarly to~\cite{Flauger2020qyi}, 
we then define our likelihood as a combination of a Gaussian and of a log-normal distribution which accounts for skewness contributions, 
giving a more accurate result for the model parameters~\cite{Bond1998qg, Sievers2002tq, Verde2003ey, Hamimeche2008ai, Flauger2020qyi}. 

The data are modeled as:
\begin{equation}
	D^{th}_{ij} (f^k_{ij}) = h^2 \, \Omega_{GW} (f^k_{ij}, \vec{\theta}_s) + h^2 \, \Omega_{n} (f^k_{ij}, \vec{\theta}_n) \;, 
\end{equation}
where $\Omega_{GW}$ is the model of the spectral shape of the signal component (here Eq.~\eqref{eq:galfit} in $\Omega$ 
units~\footnote{Notice that for this analysis, in order to work with order one numbers, rather than the parameter $A$ appearing in 
Eq.~\eqref{eq:galfit} we constrain the $\log_{10} (h^2	\, \Omega_*) = \log_{10} \left[ 4 \pi^2 A / 3 /  (H_0 /h)^2\right]$.}), 
$\vec{\theta}_s$ is the vector of parameters for the signal, $ \Omega_{n}$ is the instrument noise (\emph{i.e.} $\tilde{S}_\mathrm{instr}(f)$) 
in omega units, and $\vec{\theta}_n$ is the vector of parameters for the noise. In the following analysis we fix $\vec{\theta}_n$ at their nominal values.

Finally, we  sample the parameter space using~\texttt{Polychord}~\cite{Handley2015fda, Handley2015vkr} via its interface 
with~\texttt{Cobaya}~\cite{torrado2020cobaya}. The results of this investigation, which are analyzed and  visualized using~\texttt{GetDist}~\cite{Lewis2019xzd}, are shown in Fig.~\ref{fig:corner}. 

\bfig[h!]
\includegraphics[width=0.48\textwidth]{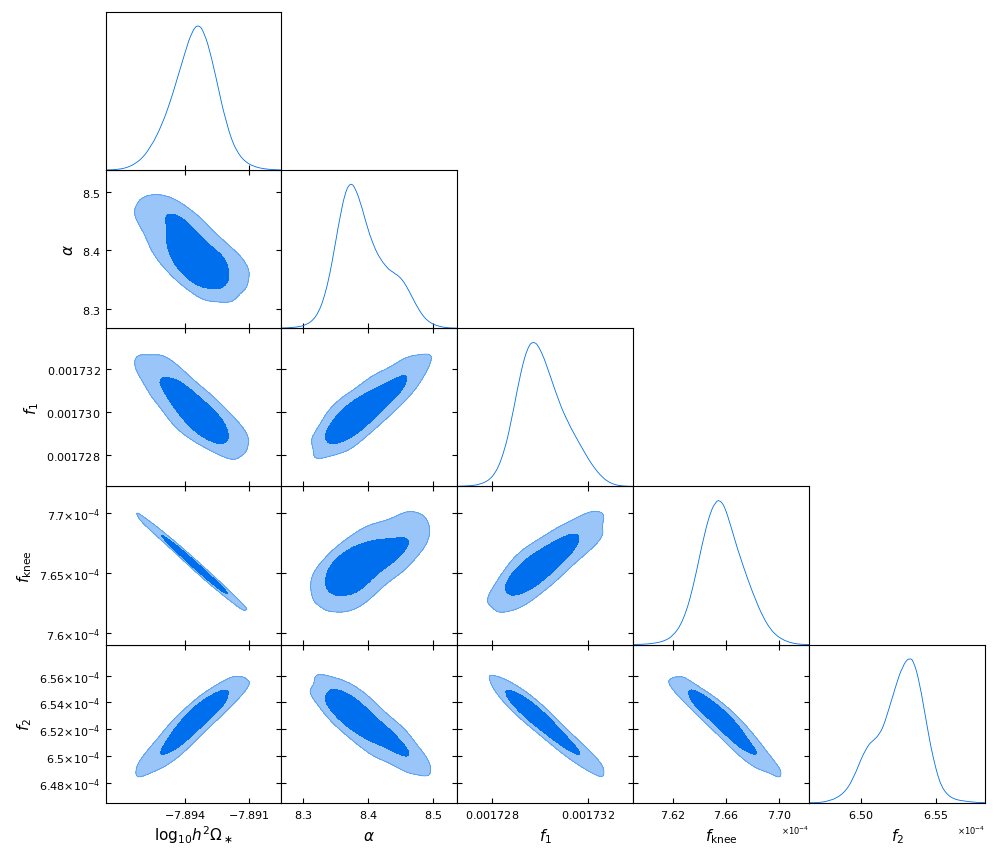}\caption{\label{fig:corner} The 2-d projections of the posterior 
distributions of the parameters of model of~\refeq{eq:galfit}, as sampled with MCMC methods (see text for details). The fit 
was performed on a data set with duration of 10 years, after subtracting the bright sources with SNR threshold $\rho_0 = 5$.}
\efig
 
To conclude this section we discuss the performance of the phenomenological model in~\refeq{eq:galfit} to fit the data. As it's 
possible to see from figure~\ref{fig:mdlfit}, the best fit for this model (shown in blue) is quite accurately matching the data 
(gray shade) for all frequencies where the confusion residual signal from the CGBs dominates over the instrumental noise (\emph{i.e.} 
for frequencies larger than ${\sim} 5 \times 10^{-4}$). On the other hand, we notice that at lower frequencies the template overestimates 
the level of confusion noise with respect to the result obtained from the simulated catalogue. In order to look for a possible 
source for this deviation, we performed further parameter estimation assuming the spectral tilt (the $-7/3$ in~\refeq{eq:galfit}) 
to be another free parameter of the model. In this case we obtained a slightly larger value for the tilt (see pink curve in 
figure~\ref{fig:mdlfit}) which leads to a better fit to the data in the low frequency band. While a detailed analysis of this 
deviation could motivate further investigations, it lies beyond the scopes of the present work and thus it is left for future studies on the topic.

\bfig[h!]
\includegraphics[width=0.4\textwidth]{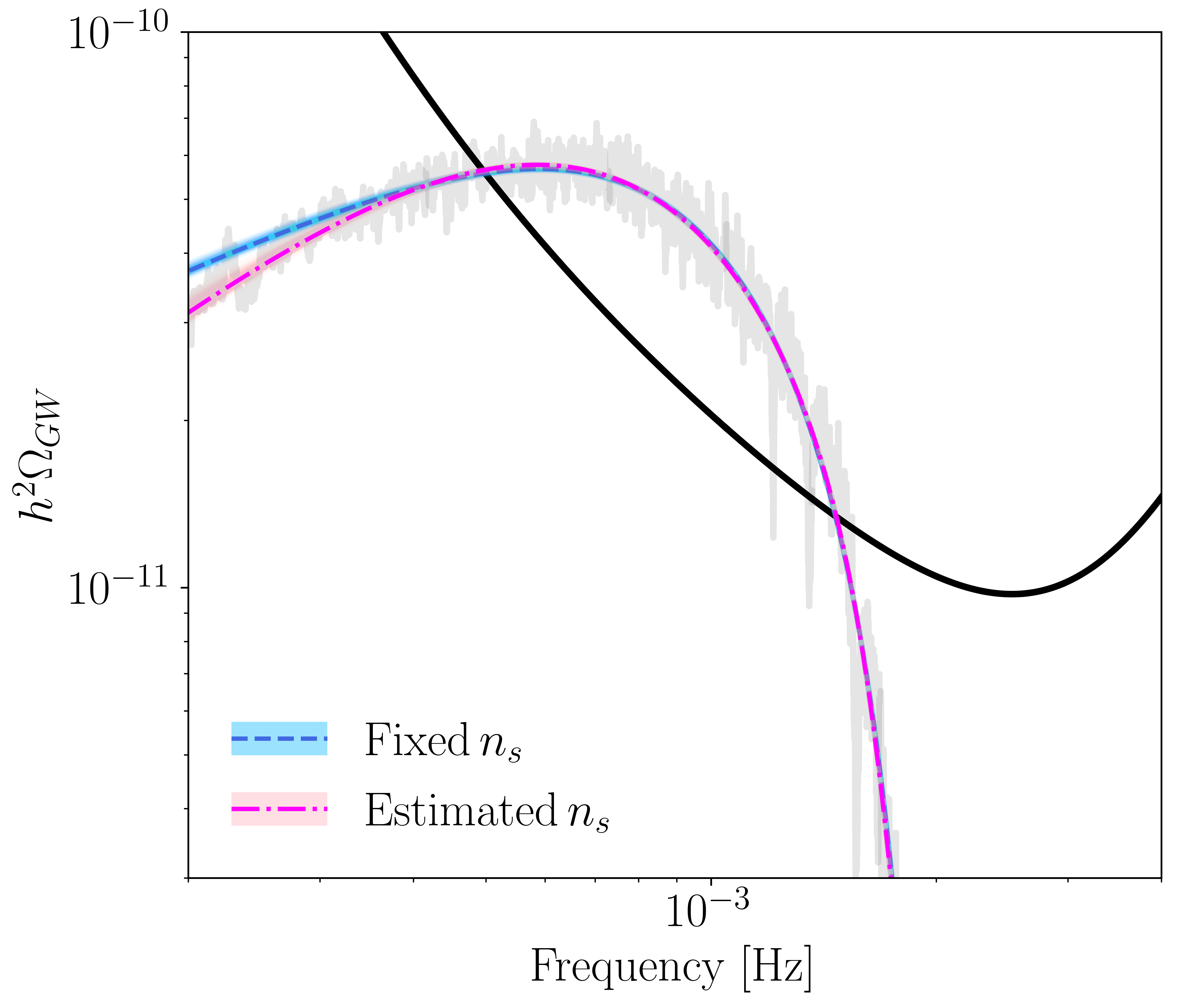}\caption{\label{fig:mdlfit} Fit of the two models we 
considered to the confusion residuals signal from CGBs. The first model, following faithfully~\refeq{eq:galfit}, 
assumes a fixed spectral index of $n_s=-7/3$ (shown in blue). The pink data refer to a model where the spectral index $n_s$ is a free 
parameter of the fit (see text for details).}
\efig

\begin{table}[b]
\caption{\label{tab:galmodel}
The parameters of the empirical model describing the confusion noise due to the unresolved CGBs signal.  We report on both methods 
of performing the data smoothing:  running mean and median}
\begin{ruledtabular}
\begin{tabular}{|c|cc|cc|}
 \multirow{3}{*}{Parameter} & \multicolumn{4}{c|}{Estimated} \\ \cline{2-5}
 & \multicolumn{2}{c}{Running mean} & \multicolumn{2}{c|}{Running median} \\ \cline{2-5}
 & $\rho_0 = 5$ & $\rho_0 = 7$ & $\rho_0 = 5$ & $\rho_0 = 7$ \\ 
\hline
 $a_1$ 				& -0.16		& -0.25		&	-0.15	& -0.15 	\\
 $a_k$ 				& -0.34		& -0.27		&	-0.34	& -0.37 	\\
 $b_1$ 				& -2.78		& -2.70		&	-2.78	& -2.72 	\\
 $b_k$ 				& -2.53		& -2.47		&	-2.55	& -2.49 	\\
 $A\times10^{-44}$	& 1.15 		& 1.14		&	1.14	& 1.15 		\\
 $f_2$ 				& 0.00059	& 0.00031	&	0.00059	& 0.00067 	\\
 $\alpha$ 			& 1.66		& 1.80		&	1.66	& 1.56 		\\
\end{tabular}
\end{ruledtabular}
\end{table}




%
%
%
%
%
%
\section{Stochastic signal due to Stellar mass Binary Black Holes
\label{sec:sobbhs}}

Since the initial successes of the ground-based GW detectors, there have been studies 
focusing on the possibility of measuring GW from SBBHs~\cite{Toubiana2020drf, Jani2019ffg, Toubiana2020cqv}. 
Based on the current understanding of the SBBH population, the expected the number of detectable sources in the LISA band ranges from $\mathcal{O}(1)$ to $\mathcal{O}(10)$.
These sources will be in the LISA band for several years before entering  and merging in the band covered by ground based interferometers~\cite{sesanasobbhs, Toubiana2020drf, Toubiana2020cqv, 101093mnrasstz897}.
LISA is sensitive to the early inspiral stage of orbital evolution of SBBHs, where the orbit shrinks very slowly and therefore most of 
the binaries stay in the observing band throughout 
the LISA observation time. 
 
Depending on the population model, the SBBH signals may fill the LISA observing band down to  ${\sim}3$mHz, thus partially overlapping with the high frequency tail of the CGB population. 
At these frequencies the CGB population transitions from being a confusion-dominated population to the regime where all sources are individually identifiable~cite{10.1093/mnras/stz2834}. 
The region of potential overlap between the CGB residual and the SBBH population reduces with the LISA mission duration, as more individual CGBs sources are being resolved and subtracted. 

The population of SBBHs observed by LISA depends on the merger rate of SBBHs in the local Universe,  
$\mathcal{R}$, expressed as the number of mergers per co-moving volume  ($\rm{Gpc}^3$) per year.
$\mathcal{R}$ is currently best estimated from the mergers detected during LIGO-Virgo observing runs, which also provide inferences on 
the mass, mass ratio and spin distribution~\cite{Abbott2020mjq}. These observations are translated into an event rate in the LISA band~\cite{sesanasobbhs} as    

\be
\frac{\mathrm{d}^2 n}{\mathrm{d}\mathcal{M} \mathrm{d}t \mathrm{d}z} = \mathcal{R} \frac{ \mathrm{d}V_c(z)} {\mathrm{d}z} \frac 1{1+z} p(\mathcal{M}),
\ee 
where $p(\mathcal{M})$ is the probability distribution function for the chirp mass
 $\mathcal{M}$ . We could include  other parameter like mass ratio, spins, etc., but for this simple example we restrict attention 
 only to the chirp mass. We have used the original prescription~\cite{sesanasobbhs} to generate a catalogue of SBBH sources emitting in 
 the LISA band. 
 
 Note that we did not use the most up-to-date population estimates \cite{ligoo3} and the mock populations used here 
 serve the illustration purposes, more elaborated analysis is underway and will be publish later~\cite{marcoccia2021}.
Instead, we have used the catalogues of SBBHs available with the first LDC data-set~\cite{ldcdata, ldcdoc}.
We simulate a dataset with duration of \tobs${\sim} 2.7$~years, using {\tt PhenomD} waveform model~\cite{phenomd}. 
We have used the same idealized assumptions about the instrument noise and subtraction fidelity as for the population of CGBs, and apply the same iterative procedure to the population of SBBHs.

Using a SNR threshold $\rho_0=15$ we do not find any detectable binaries, confirming the results of~\cite{mooregerosaklein}. 
The results for different values of $\rho_0$, are summarized in figure~\ref{fig:sobbhs}. 
It is likely that the threshold $\rho_0=15$ is an overestimate for the detection threshold, which relies on the grid-based (template bank) search which takes into account only local correlations.
On the other hand, the threshold  $\rho_0=5$ is an underestimate assuming that the volume of parameter space with high likelihood is small~\cite{ToubianaSBBH, Toubiana2020drf}.  
The results of the iterative subtraction are presented in figure~\ref{fig:sobbhs}.

\bfig[h!]
\includegraphics[width=0.48\textwidth]{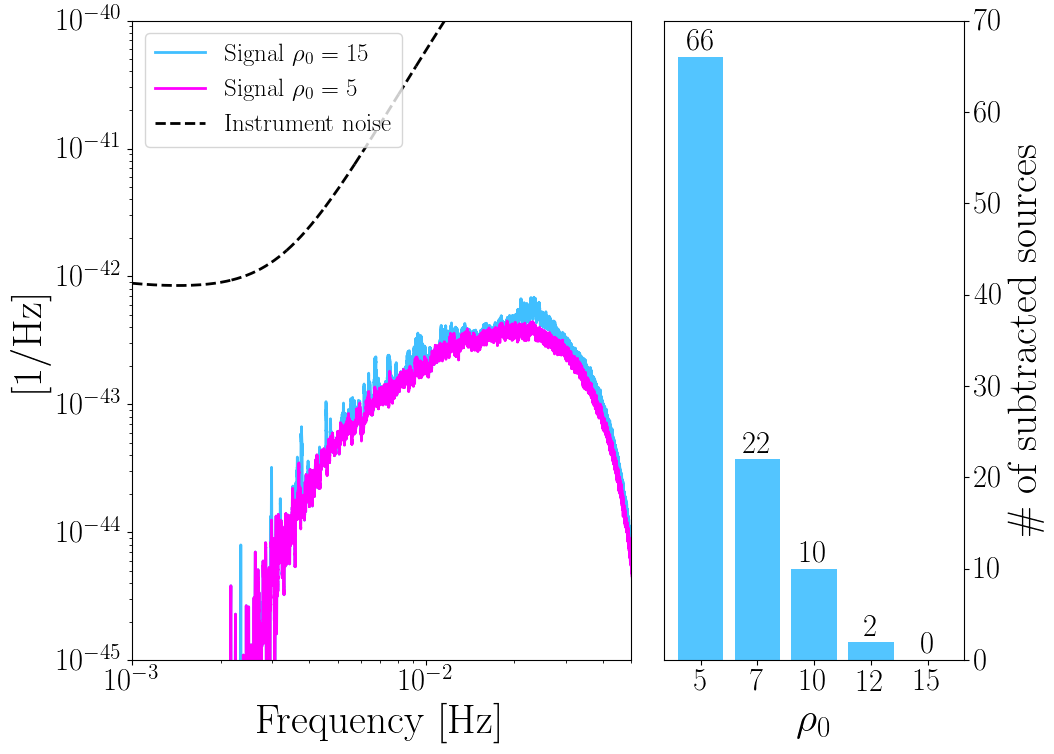}
\caption{\label{fig:sobbhs} {\em Left}: The residual confusion noise
	created by a population of SBBHs for the SNR thresholds $\rho_0 = \{5,\, 15\}$. Both curves have been estimated using a 
	running median on the data. {\em Right}: The number of resolvable sources that are 
detected and subtracted from the data for different values of SNR threshold $\rho_0$. }
\efig

%
%
%
%
%
%
\section{Combining the populations from different gravitational wave sources
\label{sec:comb}}

The same methodology and analysis principles that we used in the previous sections~(\ref{sec:gbs} and~\ref{sec:sobbhs}), 
can be followed for characterizing a foreground signal that originates from a 
combination of different types of sources. In that case, at step [c] of the algorithm described 
in section~\ref{sec:methodology}, we iterate over all types of sources and compare the individual
SNR $\rho_{i,\, \mathrm{t}}$ with $\rho_{0,\, \mathrm{t}}$, where the $\mathrm{t}$ index refers to the given
source type. Thus, the measured confusion signal would be the direct result of the superposition of the 
foreground signals from all given types of sources. This translates to 
$\tilde{S}_n(f)\equiv\tilde{S}_\mathrm{instr}(f) + \sum_\mathrm{t}\tilde{S}_\mathrm{conf,\, \mathrm{t}}(f)$,
which would yield higher levels of the unresolved foreground GW signal. Naturally, we expect that this would affect the 
source recovery and parameter estimation for the recovered sources. 
Note that this procedure does not account for the possibility of misidentifying one source type for another where the populations overlap.

For demonstration purposes, we choose to work with the two populations that we have been using so far; the CGBs, and the SBBHs. 
Using the two catalogues, we generated a LISA time series data-set with duration of \tobs$ = 2.67~\mathrm{years}$, and we applied the 
algorithm as described in section~\ref{sec:methodology}. As already mentioned, the difference here is 
that at each iteration we compute the SNR of both types of sources w.r.t the overall background noise. 
We choose an SNR threshold for each type of source, in particular $\rho_{0, \mathrm{cgb}} = 7$, and 
$\rho_{0, \mathrm{sbbh}} = \{5, \, 8,\,10,\, 12, \, 15\}$. The number of sources that were subtracted 
(i.e $\rho_{i,\, \mathrm{t}} > \rho_{0,\, \mathrm{t}}$) for each case are presented in figure~\ref{fig:combtable}. 
From our previous investigation on the same population of the SBBHs, we 
expected a similar residual confusion signal, i.e. few to no sources subtracted for high values
of $\rho_0$. This is still valid in this investigation as well. 

\bfig[h!]
\includegraphics[width=0.38\textwidth]{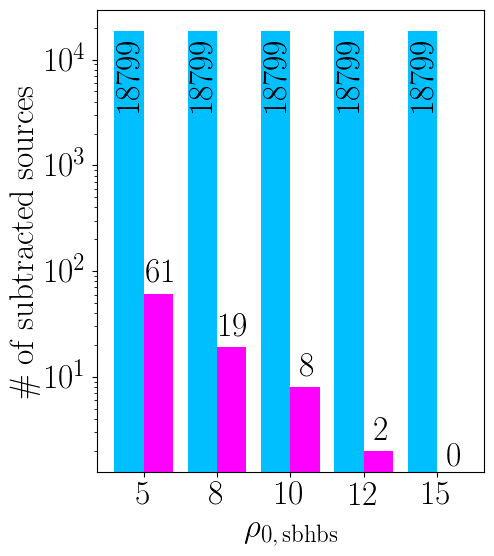}
\caption{\label{fig:combtable} The number of subtracted sources for different assumptions about the SNR threshold for 
the SBBHs, $\rho_{0, \mathrm{sbbh}}$ (represented with the pink bars). For the given population and observation duration, 
it is fairly evident, that for $\rho_{0, \mathrm{sbbh}}>15$ there are not SBBHs recovered from the data. 
The number of CGB sources that were subtracted (blue) is the same for all cases, indicating that the total 
SBBH residual confusion signal is not strong enough to affect their recovery. Here we assumed $\rho_{0, \mathrm{cgb}} = 7$. }
\efig
\begin{figure*}[ht]
\includegraphics[width=0.75\textwidth]{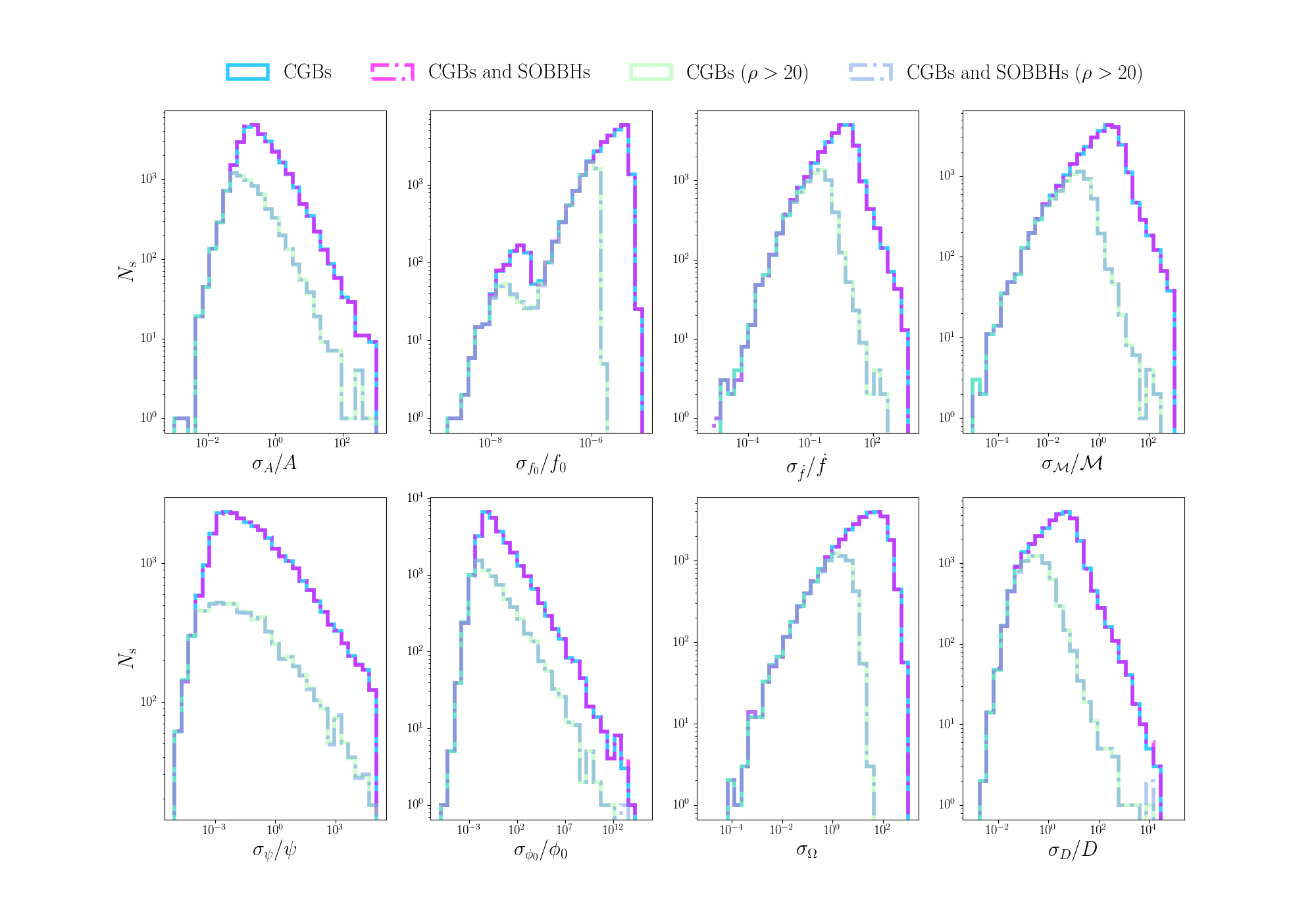}
\caption{\label{fig:comb1} Histograms of the recovered errors for the parameters of the subtracted CGBs. The optimal case, where 
there is only CGBs in the data is depicted with blue, while the relative errors of the same population with the presence of SBBHs 
is shown in pink color. The same type of histograms are plotted for the brighter ($\rho>20$) sources in green and light blue 
respectively.} 
\end{figure*}
Concerning the CGBs, we notice that almost the same number of sources are recovered 
as in section~\ref{sec:gbs}. This was also expected, due to the very low residual confusion signal of the SBBH 
component of the data, which does not impact significantly the overall measured confusion signal $\tilde{S}_n(f)$. 
But even in this case, we would expect to measure the impact of multiple components of stochastic 
signals to the recovered errors of the parameters. As described in section~\ref{sec:methodology}, at the end of each run, 
we make use of a Fisher Matrix formalism in order to get an estimate of the CGB parameters errors of the 
recovered sources. The comparison of the relative errors of the parameters compared to the ideal case of a single type of 
population is shown in figure~\ref{fig:comb1}. It is fairly evident, that indeed the precision of the CGBs 
that are recovered is minimally affected by the presence of a foreground signal generated by SBBHs, as described in~\cite{sesanasobbhs}. 
The reason lies in the characteristics of the two underlying populations. First, as we have already seen, the confusion signal 
due to the SBBHs population is small relative to the instrument noise, and thus does not significantly impact the precision with which we estimate the parameters 
of the CGBs. In addition the two populations used in this study, after subtraction of the ``loudest'' sources, only
minimally overlap to 2-3~$\mathrm{mHz}$ frequency regime.

%
%
%
%
%
%
\section{Discussion
\label{sec:conclusions}}

We have developed a methodology to perform estimates of the residual confusion signals, originating from different
types of binaries sources, as measured by LISA. We have expanded on the methodology presented in~\cite{Timpano:2005gm}, which
is based on the assumption that sources with SNR above the detection threshold will be characterized and subtracted from the 
data. We have applied this methodology, as a demonstration, to data sets containing either CGBs or SBBHs.
For the CGBs, we have updated the existing models that describe the confusion signal using up-to-date LISA 
characteristics and noise models. For the case of the 
SBBH signal, we have confirmed the findings or~\cite{mooregerosaklein}, which indicated that the population 
model of~\cite{sesanasobbhs} would yield events with quite low SNR making the detection, characterization, and subtraction of such sources challenging.

Furthermore, we have have demonstrated that one can use the same methodology to characterize the confusion signal that
is generated from different types of source populations. To do that, we have combined the components of the GW signatures of both
the CGBs, and the SBBHs that were used in the previous investigations. 
The results showed that for the given observation duration, the resulting confusion signal had an impact on the 
recovered sources of the weaker component, i.e. the one originating from the SBBHs. The number of 
recovered CGBs was not affected. Furthermore, for this particular case we tested, the accuracy of the 
parameters of the recovered CGBs was also not significantly affected. This was verified by computing the
errors of the parameters of each binary with respect to the resulting confusion noise, using a Fisher Matrix approach. The
methodology is efficient and computationally inexpensive, since the computational bottleneck was found to be just the 
template calculation. 

However, one should mention that this approach begins with idealized assumptions about the instrument noise.
Indeed, the noise is assumed to be Gaussian and stationary for the duration of the measurement, while the 
data quality is assumed optimal, i.e. uninterrupted and without any noise transients. At the same time, we have 
adopted a SNR criterion for defining the detection thresholds of the given sources. While this is proven to be a very useful 
tool for our purposes, it ignores realistic data analysis challenges, such as complications due to the 
overlap of multiple signals. Moreover, the source subtraction is assumed to be optimal as well, which means that 
the residuals are considered to be ideal. Even so, we expect the resulting characterization of confusion signal to
be an optimistic case of analyses such as~\cite{tysonpipe}. Search methods, such as the ones developed in~\cite{tysonpipe},
are capable of searching and subtracting candidate sources by utilizing the Bayesian evidence ratio (encapsulated 
in the algorithm formalism), which is a statistically more appropriate method of analysis.

Finally, this work was developed with the aim of creating a tool to characterize the foreground signal, in terms 
of spectral shape and its evolution with the LISA mission duration, for any given potential binary population that 
could contribute to it. In that sense, it presents as a useful, fast, and practical tool for zeroeth order approximation
of unresolvable astrophysical foreground signals measured by LISA.

\begin{acknowledgments}
The authors would like to acknowledge the work of the LDC group. For this study, both the LDC software and data sets were used. 
The work of M.P. was supported by STFC grants ST/P000762/1 and ST/T000791/1. M.P. acknowledges support by the European Union's 
Horizon 2020 Research Council grant 724659 MassiveCosmo ERC-2016-COG. NJC appreciates the support of NASA LISA foundation Science Grant 80NSSC19K0320
\end{acknowledgments}


\begin{thebibliography}{78}
\expandafter\ifx\csname natexlab\endcsname\relax\def\natexlab#1{#1}\fi
\expandafter\ifx\csname bibnamefont\endcsname\relax
  \def\bibnamefont#1{#1}\fi
\expandafter\ifx\csname bibfnamefont\endcsname\relax
  \def\bibfnamefont#1{#1}\fi
\expandafter\ifx\csname citenamefont\endcsname\relax
  \def\citenamefont#1{#1}\fi
\expandafter\ifx\csname url\endcsname\relax
  \def\url#1{\texttt{#1}}\fi
\expandafter\ifx\csname urlprefix\endcsname\relax\def\urlprefix{URL }\fi
\providecommand{\bibinfo}[2]{#2}
\providecommand{\eprint}[2][]{\url{#2}}

\bibitem[{\citenamefont{{Amaro-Seoane} et~al.}(2017)}]{2017arXiv170200786A}
\bibinfo{author}{\bibfnamefont{P.}~\bibnamefont{{Amaro-Seoane}}}
  \bibnamefont{et~al.}, \bibinfo{journal}{ArXiv e-prints}
  (\bibinfo{year}{2017}), \eprint{1702.00786}.

\bibitem[{\citenamefont{Baker et~al.}(2006)\citenamefont{Baker, Centrella,
  Choi, Koppitz, and van Meter}}]{PhysRevLett.96.111102}
\bibinfo{author}{\bibfnamefont{J.~G.} \bibnamefont{Baker}},
  \bibinfo{author}{\bibfnamefont{J.}~\bibnamefont{Centrella}},
  \bibinfo{author}{\bibfnamefont{D.-I.} \bibnamefont{Choi}},
  \bibinfo{author}{\bibfnamefont{M.}~\bibnamefont{Koppitz}}, \bibnamefont{and}
  \bibinfo{author}{\bibfnamefont{J.}~\bibnamefont{van Meter}},
  \bibinfo{journal}{Phys. Rev. Lett.} \textbf{\bibinfo{volume}{96}},
  \bibinfo{pages}{111102} (\bibinfo{year}{2006}),
  \urlprefix\url{https://link.aps.org/doi/10.1103/PhysRevLett.96.111102}.

\bibitem[{\citenamefont{Berti}(2006)}]{berti2006}
\bibinfo{author}{\bibfnamefont{E.}~\bibnamefont{Berti}},
  \bibinfo{journal}{Classical and Quantum Gravity}
  \textbf{\bibinfo{volume}{23}} (\bibinfo{year}{2006}).

\bibitem[{\citenamefont{Abbott et~al.}(2018)}]{ligocatalogue}
\bibinfo{author}{\bibfnamefont{B.~P.} \bibnamefont{Abbott}}
  \bibnamefont{et~al.} (\bibinfo{collaboration}{LIGO Scientific, Virgo})
  (\bibinfo{year}{2018}), \eprint{1811.12907}.

\bibitem[{\citenamefont{Abbott et~al.}(2017)}]{PhysRevLett.118.121101}
\bibinfo{author}{\bibfnamefont{B.~P.} \bibnamefont{Abbott}}
  \bibnamefont{et~al.} (\bibinfo{collaboration}{LIGO Scientific Collaboration
  and Virgo Collaboration}), \bibinfo{journal}{Phys. Rev. Lett.}
  \textbf{\bibinfo{volume}{118}}, \bibinfo{pages}{121101}
  (\bibinfo{year}{2017}),
  \urlprefix\url{https://link.aps.org/doi/10.1103/PhysRevLett.118.121101}.

\bibitem[{\citenamefont{Renzini and Contaldi}(2019)}]{PhysRevLett.122.081102}
\bibinfo{author}{\bibfnamefont{A.~I.} \bibnamefont{Renzini}} \bibnamefont{and}
  \bibinfo{author}{\bibfnamefont{C.~R.} \bibnamefont{Contaldi}},
  \bibinfo{journal}{Phys. Rev. Lett.} \textbf{\bibinfo{volume}{122}},
  \bibinfo{pages}{081102} (\bibinfo{year}{2019}),
  \urlprefix\url{https://link.aps.org/doi/10.1103/PhysRevLett.122.081102}.

\bibitem[{\citenamefont{Aasi et~al.}(2015)}]{PhysRevD.91.022003}
\bibinfo{author}{\bibfnamefont{J.}~\bibnamefont{Aasi}} \bibnamefont{et~al.}
  (\bibinfo{collaboration}{LIGO Scientific Collaboration and Virgo
  Collaboration}), \bibinfo{journal}{Phys. Rev. D}
  \textbf{\bibinfo{volume}{91}}, \bibinfo{pages}{022003}
  (\bibinfo{year}{2015}),
  \urlprefix\url{https://link.aps.org/doi/10.1103/PhysRevD.91.022003}.

\bibitem[{\citenamefont{Sesana}(2016)}]{sesanasobbhs}
\bibinfo{author}{\bibfnamefont{A.}~\bibnamefont{Sesana}},
  \bibinfo{journal}{Phys. Rev. Lett.} \textbf{\bibinfo{volume}{116}},
  \bibinfo{pages}{231102} (\bibinfo{year}{2016}),
  \urlprefix\url{https://link.aps.org/doi/10.1103/PhysRevLett.116.231102}.

\bibitem[{\citenamefont{Abbott et~al.}(2020{\natexlab{a}})}]{Abbott2020mjq}
\bibinfo{author}{\bibfnamefont{R.}~\bibnamefont{Abbott}} \bibnamefont{et~al.}
  (\bibinfo{collaboration}{LIGO Scientific, Virgo}),
  \bibinfo{journal}{Astrophys. J. Lett.} \textbf{\bibinfo{volume}{900}},
  \bibinfo{pages}{L13} (\bibinfo{year}{2020}{\natexlab{a}}),
  \eprint{2009.01190}.

\bibitem[{\citenamefont{Hiscock et~al.}(2000)\citenamefont{Hiscock, Larson,
  Routzahn, and Kulick}}]{Hiscock_2000}
\bibinfo{author}{\bibfnamefont{W.~A.} \bibnamefont{Hiscock}},
  \bibinfo{author}{\bibfnamefont{S.~L.} \bibnamefont{Larson}},
  \bibinfo{author}{\bibfnamefont{J.~R.} \bibnamefont{Routzahn}},
  \bibnamefont{and} \bibinfo{author}{\bibfnamefont{B.}~\bibnamefont{Kulick}},
  \bibinfo{journal}{The Astrophysical Journal} \textbf{\bibinfo{volume}{540}},
  \bibinfo{pages}{L5} (\bibinfo{year}{2000}),
  \urlprefix\url{https://doi.org/10.1086%2F312867}.

\bibitem[{\citenamefont{Kalogera et~al.}(2001)\citenamefont{Kalogera, Narayan,
  Spergel, and Taylor}}]{Kalogera_2001}
\bibinfo{author}{\bibfnamefont{V.}~\bibnamefont{Kalogera}},
  \bibinfo{author}{\bibfnamefont{R.}~\bibnamefont{Narayan}},
  \bibinfo{author}{\bibfnamefont{D.~N.} \bibnamefont{Spergel}},
  \bibnamefont{and} \bibinfo{author}{\bibfnamefont{J.~H.}
  \bibnamefont{Taylor}}, \bibinfo{journal}{The Astrophysical Journal}
  \textbf{\bibinfo{volume}{556}}, \bibinfo{pages}{340} (\bibinfo{year}{2001}),
  \urlprefix\url{https://doi.org/10.1086%2F321583}.

\bibitem[{\citenamefont{Nelemans et~al.}(2001)\citenamefont{Nelemans,
  Yungelson, and Portegies~Zwart}}]{Nelemans2001hp}
\bibinfo{author}{\bibfnamefont{G.}~\bibnamefont{Nelemans}},
  \bibinfo{author}{\bibfnamefont{L.}~\bibnamefont{Yungelson}},
  \bibnamefont{and} \bibinfo{author}{\bibfnamefont{S.~F.}
  \bibnamefont{Portegies~Zwart}}, \bibinfo{journal}{Astron. Astrophys.}
  \textbf{\bibinfo{volume}{375}}, \bibinfo{pages}{890} (\bibinfo{year}{2001}),
  \eprint{astro-ph/0105221}.

\bibitem[{\citenamefont{Marsh}(2011)}]{Marsh_2011}
\bibinfo{author}{\bibfnamefont{T.~R.} \bibnamefont{Marsh}},
  \bibinfo{journal}{Classical and Quantum Gravity}
  \textbf{\bibinfo{volume}{28}}, \bibinfo{pages}{094019}
  (\bibinfo{year}{2011}),
  \urlprefix\url{https://doi.org/10.1088%2F0264-9381%2F28%2F9%2F094019}.

\bibitem[{\citenamefont{Berry et~al.}(2016)\citenamefont{Berry, Cole,
  Ca\~nizares, and Gair}}]{PhysRevD.94.124042}
\bibinfo{author}{\bibfnamefont{C.~P.~L.} \bibnamefont{Berry}},
  \bibinfo{author}{\bibfnamefont{R.~H.} \bibnamefont{Cole}},
  \bibinfo{author}{\bibfnamefont{P.}~\bibnamefont{Ca\~nizares}},
  \bibnamefont{and} \bibinfo{author}{\bibfnamefont{J.~R.} \bibnamefont{Gair}},
  \bibinfo{journal}{Phys. Rev. D} \textbf{\bibinfo{volume}{94}},
  \bibinfo{pages}{124042} (\bibinfo{year}{2016}),
  \urlprefix\url{https://link.aps.org/doi/10.1103/PhysRevD.94.124042}.

\bibitem[{\citenamefont{Wang et~al.}(2012)\citenamefont{Wang, Shang, and
  Babak}}]{PhysRevD.86.104050}
\bibinfo{author}{\bibfnamefont{Y.}~\bibnamefont{Wang}},
  \bibinfo{author}{\bibfnamefont{Y.}~\bibnamefont{Shang}}, \bibnamefont{and}
  \bibinfo{author}{\bibfnamefont{S.}~\bibnamefont{Babak}},
  \bibinfo{journal}{Phys. Rev. D} \textbf{\bibinfo{volume}{86}},
  \bibinfo{pages}{104050} (\bibinfo{year}{2012}),
  \urlprefix\url{https://link.aps.org/doi/10.1103/PhysRevD.86.104050}.

\bibitem[{\citenamefont{Chua and Gair}(2015)}]{Chua_2015}
\bibinfo{author}{\bibfnamefont{A.~J.~K.} \bibnamefont{Chua}} \bibnamefont{and}
  \bibinfo{author}{\bibfnamefont{J.~R.} \bibnamefont{Gair}},
  \bibinfo{journal}{Classical and Quantum Gravity}
  \textbf{\bibinfo{volume}{32}}, \bibinfo{pages}{232002}
  (\bibinfo{year}{2015}),
  \urlprefix\url{https://doi.org/10.1088%2F0264-9381%2F32%2F23%2F232002}.

\bibitem[{\citenamefont{Barack and Cutler}(2005)}]{Barack2005aj}
\bibinfo{author}{\bibfnamefont{L.}~\bibnamefont{Barack}} \bibnamefont{and}
  \bibinfo{author}{\bibfnamefont{C.}~\bibnamefont{Cutler}}, in
  \emph{\bibinfo{booktitle}{{Laser interferometer space antenna. Proceedings,
  5th International LISA Symposium and 38th ESLAB Symposium, Noordwijk,
  Netherlands, July 12-15, 2004}}} (\bibinfo{year}{2005}),
  \eprint{gr-qc/0502052}.

\bibitem[{\citenamefont{Caprini and Figueroa}(2018)}]{Caprini_2018}
\bibinfo{author}{\bibfnamefont{C.}~\bibnamefont{Caprini}} \bibnamefont{and}
  \bibinfo{author}{\bibfnamefont{D.~G.} \bibnamefont{Figueroa}},
  \bibinfo{journal}{Classical and Quantum Gravity}
  \textbf{\bibinfo{volume}{35}}, \bibinfo{pages}{163001}
  (\bibinfo{year}{2018}),
  \urlprefix\url{https://doi.org/10.1088%2F1361-6382%2Faac608}.

\bibitem[{\citenamefont{Regimbau and Hughes}(2009)}]{PhysRevD.79.062002}
\bibinfo{author}{\bibfnamefont{T.}~\bibnamefont{Regimbau}} \bibnamefont{and}
  \bibinfo{author}{\bibfnamefont{S.~A.} \bibnamefont{Hughes}},
  \bibinfo{journal}{Phys. Rev. D} \textbf{\bibinfo{volume}{79}},
  \bibinfo{pages}{062002} (\bibinfo{year}{2009}),
  \urlprefix\url{https://link.aps.org/doi/10.1103/PhysRevD.79.062002}.

\bibitem[{\citenamefont{Robson and Cornish}(2017)}]{Robson_2017}
\bibinfo{author}{\bibfnamefont{T.}~\bibnamefont{Robson}} \bibnamefont{and}
  \bibinfo{author}{\bibfnamefont{N.}~\bibnamefont{Cornish}},
  \bibinfo{journal}{Classical and Quantum Gravity}
  \textbf{\bibinfo{volume}{34}}, \bibinfo{pages}{244002}
  (\bibinfo{year}{2017}),
  \urlprefix\url{https://doi.org/10.1088%2F1361-6382%2Faa9601}.

\bibitem[{\citenamefont{{Bender} and {Hils}}(1997)}]{1997CQGra..14.1439B}
\bibinfo{author}{\bibfnamefont{P.~L.} \bibnamefont{{Bender}}} \bibnamefont{and}
  \bibinfo{author}{\bibfnamefont{D.}~\bibnamefont{{Hils}}},
  \bibinfo{journal}{Classical and Quantum Gravity}
  \textbf{\bibinfo{volume}{14}}, \bibinfo{pages}{1439} (\bibinfo{year}{1997}).

\bibitem[{\citenamefont{{Barack} and {Cutler}}(2004)}]{2004PhRvD..70l2002B}
\bibinfo{author}{\bibfnamefont{L.}~\bibnamefont{{Barack}}} \bibnamefont{and}
  \bibinfo{author}{\bibfnamefont{C.}~\bibnamefont{{Cutler}}},
  \bibinfo{journal}{\prd} \textbf{\bibinfo{volume}{70}}, \bibinfo{eid}{122002}
  (\bibinfo{year}{2004}), \eprint{gr-qc/0409010}.

\bibitem[{\citenamefont{Timpano et~al.}(2006)\citenamefont{Timpano, Rubbo, and
  Cornish}}]{Timpano:2005gm}
\bibinfo{author}{\bibfnamefont{S.~E.} \bibnamefont{Timpano}},
  \bibinfo{author}{\bibfnamefont{L.~J.} \bibnamefont{Rubbo}}, \bibnamefont{and}
  \bibinfo{author}{\bibfnamefont{N.~J.} \bibnamefont{Cornish}},
  \bibinfo{journal}{Phys. Rev. D} \textbf{\bibinfo{volume}{73}},
  \bibinfo{pages}{122001} (\bibinfo{year}{2006}), \eprint{gr-qc/0504071}.

\bibitem[{\citenamefont{{LISA Science Study Team}}(2018)}]{scird}
\bibinfo{author}{\bibnamefont{{LISA Science Study Team}}}, \bibinfo{type}{Tech.
  Rep.} \bibinfo{number}{{ESA-L3-EST-SCI-RS-001}}, \bibinfo{institution}{ESA}
  (\bibinfo{year}{2018}),
  \urlprefix\url{https://www.cosmos.esa.int/documents/678316/1700384/SciRD.pdf}.

\bibitem[{\citenamefont{Stroeer and Vecchio}(2006)}]{Stroeer_2006}
\bibinfo{author}{\bibfnamefont{A.}~\bibnamefont{Stroeer}} \bibnamefont{and}
  \bibinfo{author}{\bibfnamefont{A.}~\bibnamefont{Vecchio}},
  \bibinfo{journal}{Classical and Quantum Gravity}
  \textbf{\bibinfo{volume}{23}}, \bibinfo{pages}{S809} (\bibinfo{year}{2006}),
  \urlprefix\url{https://doi.org/10.1088%2F0264-9381%2F23%2F19%2Fs19}.

\bibitem[{\citenamefont{Kupfer et~al.}(2018)\citenamefont{Kupfer, Korol, Shah,
  Nelemans, Marsh, Ramsay, Groot, Steeghs, and Rossi}}]{10.1093/mnras/sty1545}
\bibinfo{author}{\bibfnamefont{T.}~\bibnamefont{Kupfer}},
  \bibinfo{author}{\bibfnamefont{V.}~\bibnamefont{Korol}},
  \bibinfo{author}{\bibfnamefont{S.}~\bibnamefont{Shah}},
  \bibinfo{author}{\bibfnamefont{G.}~\bibnamefont{Nelemans}},
  \bibinfo{author}{\bibfnamefont{T.~R.} \bibnamefont{Marsh}},
  \bibinfo{author}{\bibfnamefont{G.}~\bibnamefont{Ramsay}},
  \bibinfo{author}{\bibfnamefont{P.~J.} \bibnamefont{Groot}},
  \bibinfo{author}{\bibfnamefont{D.~T.~H.} \bibnamefont{Steeghs}},
  \bibnamefont{and} \bibinfo{author}{\bibfnamefont{E.~M.} \bibnamefont{Rossi}},
  \bibinfo{journal}{Monthly Notices of the Royal Astronomical Society}
  \textbf{\bibinfo{volume}{480}}, \bibinfo{pages}{302} (\bibinfo{year}{2018}),
  ISSN \bibinfo{issn}{0035-8711},
  \eprint{https://academic.oup.com/mnras/article-pdf/480/1/302/25244351/sty1545.pdf},
  \urlprefix\url{https://doi.org/10.1093/mnras/sty1545}.

\bibitem[{\citenamefont{Nelemans et~al.}(2004)\citenamefont{Nelemans,
  Yungelson, and Portegies~Zwart}}]{Nelemans_2004}
\bibinfo{author}{\bibfnamefont{G.}~\bibnamefont{Nelemans}},
  \bibinfo{author}{\bibfnamefont{L.~R.} \bibnamefont{Yungelson}},
  \bibnamefont{and} \bibinfo{author}{\bibfnamefont{S.~F.}
  \bibnamefont{Portegies~Zwart}}, \bibinfo{journal}{Monthly Notices of the
  Royal Astronomical Society} \textbf{\bibinfo{volume}{349}},
  \bibinfo{pages}{181} (\bibinfo{year}{2004}), ISSN \bibinfo{issn}{0035-8711},
  \eprint{https://academic.oup.com/mnras/article-pdf/349/1/181/11180320/349-1-181.pdf},
  \urlprefix\url{https://doi.org/10.1111/j.1365-2966.2004.07479.x}.

\bibitem[{\citenamefont{Korol et~al.}(2018)\citenamefont{Korol, Koop, and
  Rossi}}]{Korol_2018}
\bibinfo{author}{\bibfnamefont{V.}~\bibnamefont{Korol}},
  \bibinfo{author}{\bibfnamefont{O.}~\bibnamefont{Koop}}, \bibnamefont{and}
  \bibinfo{author}{\bibfnamefont{E.~M.} \bibnamefont{Rossi}},
  \bibinfo{journal}{The Astrophysical Journal} \textbf{\bibinfo{volume}{866}},
  \bibinfo{pages}{L20} (\bibinfo{year}{2018}),
  \urlprefix\url{https://doi.org/10.3847%2F2041-8213%2Faae587}.

\bibitem[{\citenamefont{Korol et~al.}(2017)\citenamefont{Korol, Rossi, Groot,
  Nelemans, Toonen, and Brown}}]{Korol_2017}
\bibinfo{author}{\bibfnamefont{V.}~\bibnamefont{Korol}},
  \bibinfo{author}{\bibfnamefont{E.~M.} \bibnamefont{Rossi}},
  \bibinfo{author}{\bibfnamefont{P.~J.} \bibnamefont{Groot}},
  \bibinfo{author}{\bibfnamefont{G.}~\bibnamefont{Nelemans}},
  \bibinfo{author}{\bibfnamefont{S.}~\bibnamefont{Toonen}}, \bibnamefont{and}
  \bibinfo{author}{\bibfnamefont{A.~G.~A.} \bibnamefont{Brown}},
  \bibinfo{journal}{Monthly Notices of the Royal Astronomical Society}
  \textbf{\bibinfo{volume}{470}}, \bibinfo{pages}{1894} (\bibinfo{year}{2017}),
  ISSN \bibinfo{issn}{0035-8711},
  \eprint{https://academic.oup.com/mnras/article-pdf/470/2/1894/18139327/stx1285.pdf},
  \urlprefix\url{https://doi.org/10.1093/mnras/stx1285}.

\bibitem[{\citenamefont{Abbott et~al.}(2020{\natexlab{b}})}]{ligoo3}
\bibinfo{author}{\bibfnamefont{R.}~\bibnamefont{Abbott}} \bibnamefont{et~al.}
  (\bibinfo{collaboration}{LIGO Scientific, Virgo})
  (\bibinfo{year}{2020}{\natexlab{b}}), \eprint{2010.14527}.

\bibitem[{\citenamefont{Moore et~al.}(2019)\citenamefont{Moore, Gerosa, and
  Klein}}]{mooregerosaklein}
\bibinfo{author}{\bibfnamefont{C.~J.} \bibnamefont{Moore}},
  \bibinfo{author}{\bibfnamefont{D.}~\bibnamefont{Gerosa}}, \bibnamefont{and}
  \bibinfo{author}{\bibfnamefont{A.}~\bibnamefont{Klein}},
  \bibinfo{journal}{Monthly Notices of the Royal Astronomical Society: Letters}
  \textbf{\bibinfo{volume}{488}}, \bibinfo{pages}{L94} (\bibinfo{year}{2019}),
  ISSN \bibinfo{issn}{1745-3925},
  \eprint{https://academic.oup.com/mnrasl/article-pdf/488/1/L94/28944022/slz104.pdf},
  \urlprefix\url{https://doi.org/10.1093/mnrasl/slz104}.

\bibitem[{\citenamefont{Abbott et~al.}(2007)}]{Abbott_2007}
\bibinfo{author}{\bibfnamefont{B.}~\bibnamefont{Abbott}} \bibnamefont{et~al.},
  \bibinfo{journal}{The Astrophysical Journal} \textbf{\bibinfo{volume}{659}},
  \bibinfo{pages}{918} (\bibinfo{year}{2007}),
  \urlprefix\url{https://doi.org/10.1086%2F511329}.

\bibitem[{\citenamefont{Abbott et~al.}(2016)}]{PhysRevD.93.042005}
\bibinfo{author}{\bibfnamefont{B.~P.} \bibnamefont{Abbott}}
  \bibnamefont{et~al.} (\bibinfo{collaboration}{The LIGO Scientific
  Collaboration and the Virgo Collaboration}), \bibinfo{journal}{Phys. Rev. D}
  \textbf{\bibinfo{volume}{93}}, \bibinfo{pages}{042005}
  (\bibinfo{year}{2016}),
  \urlprefix\url{https://link.aps.org/doi/10.1103/PhysRevD.93.042005}.

\bibitem[{\citenamefont{Arzoumanian et~al.}(2016)}]{Arzoumanian_2016}
\bibinfo{author}{\bibfnamefont{Z.}~\bibnamefont{Arzoumanian}}
  \bibnamefont{et~al.}, \bibinfo{journal}{The Astrophysical Journal}
  \textbf{\bibinfo{volume}{821}}, \bibinfo{pages}{13} (\bibinfo{year}{2016}),
  \urlprefix\url{https://doi.org/10.3847%2F0004-637x%2F821%2F1%2F13}.

\bibitem[{\citenamefont{Arzoumanian et~al.}(2018)}]{Arzoumanian_2018}
\bibinfo{author}{\bibfnamefont{Z.}~\bibnamefont{Arzoumanian}}
  \bibnamefont{et~al.}, \bibinfo{journal}{The Astrophysical Journal}
  \textbf{\bibinfo{volume}{859}}, \bibinfo{pages}{47} (\bibinfo{year}{2018}),
  \urlprefix\url{https://doi.org/10.3847%2F1538-4357%2Faabd3b}.

\bibitem[{\citenamefont{Gair et~al.}(2015)\citenamefont{Gair, Romano, and
  Taylor}}]{PhysRevD.92.102003}
\bibinfo{author}{\bibfnamefont{J.~R.} \bibnamefont{Gair}},
  \bibinfo{author}{\bibfnamefont{J.~D.} \bibnamefont{Romano}},
  \bibnamefont{and} \bibinfo{author}{\bibfnamefont{S.~R.}
  \bibnamefont{Taylor}}, \bibinfo{journal}{Phys. Rev. D}
  \textbf{\bibinfo{volume}{92}}, \bibinfo{pages}{102003}
  (\bibinfo{year}{2015}),
  \urlprefix\url{https://link.aps.org/doi/10.1103/PhysRevD.92.102003}.

\bibitem[{\citenamefont{Lentati et~al.}(2015)}]{Lentati2015qwp}
\bibinfo{author}{\bibfnamefont{L.}~\bibnamefont{Lentati}} \bibnamefont{et~al.},
  \bibinfo{journal}{Monthly Notices of the Royal Astronomical Society}
  \textbf{\bibinfo{volume}{453}}, \bibinfo{pages}{2576} (\bibinfo{year}{2015}),
  ISSN \bibinfo{issn}{0035-8711},
  \eprint{https://academic.oup.com/mnras/article-pdf/453/3/2576/4900884/stv1538.pdf},
  \urlprefix\url{https://doi.org/10.1093/mnras/stv1538}.

\bibitem[{\citenamefont{Arzoumanian et~al.}(2020)}]{Nanograv12.5}
\bibinfo{author}{\bibfnamefont{Z.}~\bibnamefont{Arzoumanian}}
  \bibnamefont{et~al.} (\bibinfo{collaboration}{NANOGrav})
  (\bibinfo{year}{2020}), \eprint{2009.04496}.

\bibitem[{\citenamefont{Babak et~al.}(2017)\citenamefont{Babak, Gair, Sesana,
  Barausse, Sopuerta, Berry, Berti, Amaro-Seoane, Petiteau, and
  Klein}}]{BabakEMRIs}
\bibinfo{author}{\bibfnamefont{S.}~\bibnamefont{Babak}},
  \bibinfo{author}{\bibfnamefont{J.}~\bibnamefont{Gair}},
  \bibinfo{author}{\bibfnamefont{A.}~\bibnamefont{Sesana}},
  \bibinfo{author}{\bibfnamefont{E.}~\bibnamefont{Barausse}},
  \bibinfo{author}{\bibfnamefont{C.~F.} \bibnamefont{Sopuerta}},
  \bibinfo{author}{\bibfnamefont{C.~P.} \bibnamefont{Berry}},
  \bibinfo{author}{\bibfnamefont{E.}~\bibnamefont{Berti}},
  \bibinfo{author}{\bibfnamefont{P.}~\bibnamefont{Amaro-Seoane}},
  \bibinfo{author}{\bibfnamefont{A.}~\bibnamefont{Petiteau}}, \bibnamefont{and}
  \bibinfo{author}{\bibfnamefont{A.}~\bibnamefont{Klein}},
  \bibinfo{journal}{Phys. Rev. D} \textbf{\bibinfo{volume}{95}},
  \bibinfo{pages}{103012} (\bibinfo{year}{2017}), \eprint{1703.09722}.

\bibitem[{\citenamefont{Nissanke et~al.}(2012)\citenamefont{Nissanke,
  Vallisneri, Nelemans, and Prince}}]{Nissanke_2012}
\bibinfo{author}{\bibfnamefont{S.}~\bibnamefont{Nissanke}},
  \bibinfo{author}{\bibfnamefont{M.}~\bibnamefont{Vallisneri}},
  \bibinfo{author}{\bibfnamefont{G.}~\bibnamefont{Nelemans}}, \bibnamefont{and}
  \bibinfo{author}{\bibfnamefont{T.~A.} \bibnamefont{Prince}},
  \bibinfo{journal}{The Astrophysical Journal} \textbf{\bibinfo{volume}{758}},
  \bibinfo{pages}{131} (\bibinfo{year}{2012}),
  \urlprefix\url{https://doi.org/10.1088%2F0004-637x%2F758%2F2%2F131}.

\bibitem[{\citenamefont{Cornish and Robson}(2017)}]{Cornish2017vip}
\bibinfo{author}{\bibfnamefont{N.}~\bibnamefont{Cornish}} \bibnamefont{and}
  \bibinfo{author}{\bibfnamefont{T.}~\bibnamefont{Robson}},
  \bibinfo{journal}{J. Phys. Conf. Ser.} \textbf{\bibinfo{volume}{840}},
  \bibinfo{pages}{012024} (\bibinfo{year}{2017}), \eprint{1703.09858}.

\bibitem[{\citenamefont{Crowder and Cornish}(2007)}]{PhysRevD.75.043008}
\bibinfo{author}{\bibfnamefont{J.}~\bibnamefont{Crowder}} \bibnamefont{and}
  \bibinfo{author}{\bibfnamefont{N.~J.} \bibnamefont{Cornish}},
  \bibinfo{journal}{Phys. Rev. D} \textbf{\bibinfo{volume}{75}},
  \bibinfo{pages}{043008} (\bibinfo{year}{2007}),
  \urlprefix\url{https://link.aps.org/doi/10.1103/PhysRevD.75.043008}.

\bibitem[{\citenamefont{Tinto and Dhurandhar}(2005)}]{tdi}
\bibinfo{author}{\bibfnamefont{M.}~\bibnamefont{Tinto}} \bibnamefont{and}
  \bibinfo{author}{\bibfnamefont{S.~V.} \bibnamefont{Dhurandhar}},
  \bibinfo{journal}{Living Reviews in Relativity} \textbf{\bibinfo{volume}{8}},
  \bibinfo{pages}{4} (\bibinfo{year}{2005}), ISSN \bibinfo{issn}{1433-8351},
  \urlprefix\url{https://doi.org/10.12942/lrr-2005-4}.

\bibitem[{\citenamefont{Prince et~al.}(2002)\citenamefont{Prince, Tinto,
  Larson, and Armstrong}}]{aet}
\bibinfo{author}{\bibfnamefont{T.~A.} \bibnamefont{Prince}},
  \bibinfo{author}{\bibfnamefont{M.}~\bibnamefont{Tinto}},
  \bibinfo{author}{\bibfnamefont{S.~L.} \bibnamefont{Larson}},
  \bibnamefont{and} \bibinfo{author}{\bibfnamefont{J.~W.}
  \bibnamefont{Armstrong}}, \bibinfo{journal}{Phys. Rev. D}
  \textbf{\bibinfo{volume}{66}}, \bibinfo{pages}{122002}
  (\bibinfo{year}{2002}),
  \urlprefix\url{https://link.aps.org/doi/10.1103/PhysRevD.66.122002}.

\bibitem[{\citenamefont{Petiteau et~al.}(2008)\citenamefont{Petiteau, Auger,
  Halloin, Jeannin, Plagnol, Pireaux, Regimbau, and Vinet}}]{lisacode}
\bibinfo{author}{\bibfnamefont{A.}~\bibnamefont{Petiteau}},
  \bibinfo{author}{\bibfnamefont{G.}~\bibnamefont{Auger}},
  \bibinfo{author}{\bibfnamefont{H.}~\bibnamefont{Halloin}},
  \bibinfo{author}{\bibfnamefont{O.}~\bibnamefont{Jeannin}},
  \bibinfo{author}{\bibfnamefont{E.}~\bibnamefont{Plagnol}},
  \bibinfo{author}{\bibfnamefont{S.}~\bibnamefont{Pireaux}},
  \bibinfo{author}{\bibfnamefont{T.}~\bibnamefont{Regimbau}}, \bibnamefont{and}
  \bibinfo{author}{\bibfnamefont{J.-Y.} \bibnamefont{Vinet}},
  \bibinfo{journal}{Phys. Rev. D} \textbf{\bibinfo{volume}{77}},
  \bibinfo{pages}{023002} (\bibinfo{year}{2008}),
  \urlprefix\url{https://link.aps.org/doi/10.1103/PhysRevD.77.023002}.

\bibitem[{\citenamefont{Littenberg et~al.}(2020)\citenamefont{Littenberg,
  Cornish, Lackeos, and Robson}}]{tysonpipe}
\bibinfo{author}{\bibfnamefont{T.~B.} \bibnamefont{Littenberg}},
  \bibinfo{author}{\bibfnamefont{N.~J.} \bibnamefont{Cornish}},
  \bibinfo{author}{\bibfnamefont{K.}~\bibnamefont{Lackeos}}, \bibnamefont{and}
  \bibinfo{author}{\bibfnamefont{T.}~\bibnamefont{Robson}},
  \bibinfo{journal}{Phys. Rev. D} \textbf{\bibinfo{volume}{101}},
  \bibinfo{pages}{123021} (\bibinfo{year}{2020}),
  \urlprefix\url{https://link.aps.org/doi/10.1103/PhysRevD.101.123021}.

\bibitem[{\citenamefont{Pankow et~al.}(2018)}]{ligoglitch}
\bibinfo{author}{\bibfnamefont{C.}~\bibnamefont{Pankow}} \bibnamefont{et~al.},
  \bibinfo{journal}{Phys. Rev. D} \textbf{\bibinfo{volume}{98}},
  \bibinfo{pages}{084016} (\bibinfo{year}{2018}), \eprint{1808.03619}.

\bibitem[{\citenamefont{Robson and Cornish}(2019)}]{Robson:2018jly}
\bibinfo{author}{\bibfnamefont{T.}~\bibnamefont{Robson}} \bibnamefont{and}
  \bibinfo{author}{\bibfnamefont{N.~J.} \bibnamefont{Cornish}},
  \bibinfo{journal}{Phys. Rev. D} \textbf{\bibinfo{volume}{99}},
  \bibinfo{pages}{024019} (\bibinfo{year}{2019}), \eprint{1811.04490}.

\bibitem[{\citenamefont{Cornish et~al.}(2021)\citenamefont{Cornish, Littenberg,
  B\'ecsy, Chatziioannou, Clark, Ghonge, and Millhouse}}]{Cornish:2020dwh}
\bibinfo{author}{\bibfnamefont{N.~J.} \bibnamefont{Cornish}},
  \bibinfo{author}{\bibfnamefont{T.~B.} \bibnamefont{Littenberg}},
  \bibinfo{author}{\bibfnamefont{B.}~\bibnamefont{B\'ecsy}},
  \bibinfo{author}{\bibfnamefont{K.}~\bibnamefont{Chatziioannou}},
  \bibinfo{author}{\bibfnamefont{J.~A.} \bibnamefont{Clark}},
  \bibinfo{author}{\bibfnamefont{S.}~\bibnamefont{Ghonge}}, \bibnamefont{and}
  \bibinfo{author}{\bibfnamefont{M.}~\bibnamefont{Millhouse}},
  \bibinfo{journal}{Phys. Rev. D} \textbf{\bibinfo{volume}{103}},
  \bibinfo{pages}{044006} (\bibinfo{year}{2021}), \eprint{2011.09494}.

\bibitem[{\citenamefont{Chatziioannou et~al.}(2021)\citenamefont{Chatziioannou,
  Cornish, Wijngaarden, and Littenberg}}]{Chatziioannou:2021ezd}
\bibinfo{author}{\bibfnamefont{K.}~\bibnamefont{Chatziioannou}},
  \bibinfo{author}{\bibfnamefont{N.}~\bibnamefont{Cornish}},
  \bibinfo{author}{\bibfnamefont{M.}~\bibnamefont{Wijngaarden}},
  \bibnamefont{and} \bibinfo{author}{\bibfnamefont{T.~B.}
  \bibnamefont{Littenberg}}, \bibinfo{journal}{Phys. Rev. D}
  \textbf{\bibinfo{volume}{103}}, \bibinfo{pages}{044013}
  (\bibinfo{year}{2021}), \eprint{2101.01200}.

\bibitem[{\citenamefont{Marsat et~al.}(2020)\citenamefont{Marsat, Baker, and
  Dal~Canton}}]{MarsatBakerMBHB}
\bibinfo{author}{\bibfnamefont{S.}~\bibnamefont{Marsat}},
  \bibinfo{author}{\bibfnamefont{J.~G.} \bibnamefont{Baker}}, \bibnamefont{and}
  \bibinfo{author}{\bibfnamefont{T.}~\bibnamefont{Dal~Canton}}
  (\bibinfo{year}{2020}), \eprint{2003.00357}.

\bibitem[{\citenamefont{Cornish and Shuman}(2020)}]{NeilMBHB}
\bibinfo{author}{\bibfnamefont{N.~J.} \bibnamefont{Cornish}} \bibnamefont{and}
  \bibinfo{author}{\bibfnamefont{K.}~\bibnamefont{Shuman}},
  \bibinfo{journal}{Phys. Rev. D} \textbf{\bibinfo{volume}{101}},
  \bibinfo{pages}{124008} (\bibinfo{year}{2020}), \eprint{2005.03610}.

\bibitem[{\citenamefont{Toubiana
  et~al.}(2020{\natexlab{a}})\citenamefont{Toubiana, Marsat, Babak, Baker, and
  Dal~Canton}}]{ToubianaSBBH}
\bibinfo{author}{\bibfnamefont{A.}~\bibnamefont{Toubiana}},
  \bibinfo{author}{\bibfnamefont{S.}~\bibnamefont{Marsat}},
  \bibinfo{author}{\bibfnamefont{S.}~\bibnamefont{Babak}},
  \bibinfo{author}{\bibfnamefont{J.}~\bibnamefont{Baker}}, \bibnamefont{and}
  \bibinfo{author}{\bibfnamefont{T.}~\bibnamefont{Dal~Canton}}
  (\bibinfo{year}{2020}{\natexlab{a}}), \eprint{2007.08544}.

\bibitem[{\citenamefont{Littenberg and Yunes}(2019)}]{Littenberg_2019}
\bibinfo{author}{\bibfnamefont{T.~B.} \bibnamefont{Littenberg}}
  \bibnamefont{and} \bibinfo{author}{\bibfnamefont{N.}~\bibnamefont{Yunes}},
  \bibinfo{journal}{Classical and Quantum Gravity}
  \textbf{\bibinfo{volume}{36}}, \bibinfo{pages}{095017}
  (\bibinfo{year}{2019}), ISSN \bibinfo{issn}{1361-6382},
  \urlprefix\url{http://dx.doi.org/10.1088/1361-6382/ab0a3d}.

\bibitem[{\citenamefont{Robson et~al.}(2018)\citenamefont{Robson, Cornish,
  Tamanini, and Toonen}}]{Robson:2018svj}
\bibinfo{author}{\bibfnamefont{T.}~\bibnamefont{Robson}},
  \bibinfo{author}{\bibfnamefont{N.~J.} \bibnamefont{Cornish}},
  \bibinfo{author}{\bibfnamefont{N.}~\bibnamefont{Tamanini}}, \bibnamefont{and}
  \bibinfo{author}{\bibfnamefont{S.}~\bibnamefont{Toonen}},
  \bibinfo{journal}{Phys. Rev. D} \textbf{\bibinfo{volume}{98}},
  \bibinfo{pages}{064012} (\bibinfo{year}{2018}), \eprint{1806.00500}.

\bibitem[{\citenamefont{Danielski et~al.}(2019)\citenamefont{Danielski, Korol,
  Tamanini, and Rossi}}]{lisaexoplanets}
\bibinfo{author}{\bibfnamefont{C.}~\bibnamefont{Danielski}},
  \bibinfo{author}{\bibfnamefont{V.}~\bibnamefont{Korol}},
  \bibinfo{author}{\bibfnamefont{N.}~\bibnamefont{Tamanini}}, \bibnamefont{and}
  \bibinfo{author}{\bibfnamefont{E.}~\bibnamefont{Rossi}},
  \bibinfo{journal}{Astron. Astrophys.} \textbf{\bibinfo{volume}{632}},
  \bibinfo{pages}{A113} (\bibinfo{year}{2019}), \eprint{1910.05414}.

\bibitem[{\citenamefont{Babak and Petiteau}(2020)}]{ldcdoc}
\bibinfo{author}{\bibfnamefont{S.}~\bibnamefont{Babak}} \bibnamefont{and}
  \bibinfo{author}{\bibfnamefont{A.}~\bibnamefont{Petiteau}},
  \bibinfo{type}{Tech. Rep.} \bibinfo{number}{{LISA-LCST-SGS-MAN-002}},
  \bibinfo{institution}{APC Paris} (\bibinfo{year}{2020}),
  \urlprefix\url{https://lisa-ldc.lal.in2p3.fr/static/data/pdf/LDC-manual-002.pdf}.

\bibitem[{\citenamefont{Cornish and Littenberg}(2007)}]{gbwf}
\bibinfo{author}{\bibfnamefont{N.~J.} \bibnamefont{Cornish}} \bibnamefont{and}
  \bibinfo{author}{\bibfnamefont{T.~B.} \bibnamefont{Littenberg}},
  \bibinfo{journal}{Phys. Rev. D} \textbf{\bibinfo{volume}{76}},
  \bibinfo{pages}{083006} (\bibinfo{year}{2007}),
  \urlprefix\url{https://link.aps.org/doi/10.1103/PhysRevD.76.083006}.

\bibitem[{ldc()}]{ldcdata}
\emph{\bibinfo{title}{{LISA Data Challenges}, the {{\em Radler}} data set}},
  \bibinfo{howpublished}{\url{https://lisa-ldc.lal.in2p3.fr/ldc}},
  \bibinfo{note}{accessed: 2019-04-09}.

\bibitem[{\citenamefont{Korol et~al.}(2020)}]{Korol2020l}
\bibinfo{author}{\bibfnamefont{V.}~\bibnamefont{Korol}} \bibnamefont{et~al.},
  \bibinfo{journal}{Astron. Astrophys.} \textbf{\bibinfo{volume}{638}},
  \bibinfo{pages}{A153} (\bibinfo{year}{2020}), \eprint{2002.10462}.

\bibitem[{\citenamefont{Nelemans}(2013)}]{Nelemans2013yg}
\bibinfo{author}{\bibfnamefont{G.}~\bibnamefont{Nelemans}},
  \bibinfo{journal}{ASP Conf. Ser.} \textbf{\bibinfo{volume}{467}},
  \bibinfo{pages}{27} (\bibinfo{year}{2013}), \eprint{1302.0138}.

\bibitem[{\citenamefont{Caprini et~al.}(2019)\citenamefont{Caprini, Figueroa,
  Flauger, Nardini, Peloso, Pieroni, Ricciardone, and
  Tasinato}}]{Caprini2019pxz}
\bibinfo{author}{\bibfnamefont{C.}~\bibnamefont{Caprini}},
  \bibinfo{author}{\bibfnamefont{D.~G.} \bibnamefont{Figueroa}},
  \bibinfo{author}{\bibfnamefont{R.}~\bibnamefont{Flauger}},
  \bibinfo{author}{\bibfnamefont{G.}~\bibnamefont{Nardini}},
  \bibinfo{author}{\bibfnamefont{M.}~\bibnamefont{Peloso}},
  \bibinfo{author}{\bibfnamefont{M.}~\bibnamefont{Pieroni}},
  \bibinfo{author}{\bibfnamefont{A.}~\bibnamefont{Ricciardone}},
  \bibnamefont{and} \bibinfo{author}{\bibfnamefont{G.}~\bibnamefont{Tasinato}},
  \bibinfo{journal}{JCAP} \textbf{\bibinfo{volume}{11}}, \bibinfo{pages}{017}
  (\bibinfo{year}{2019}), \eprint{1906.09244}.

\bibitem[{\citenamefont{Pieroni and Barausse}(2020)}]{Pieroni2020rob}
\bibinfo{author}{\bibfnamefont{M.}~\bibnamefont{Pieroni}} \bibnamefont{and}
  \bibinfo{author}{\bibfnamefont{E.}~\bibnamefont{Barausse}},
  \bibinfo{journal}{JCAP} \textbf{\bibinfo{volume}{07}}, \bibinfo{pages}{021}
  (\bibinfo{year}{2020}), \bibinfo{note}{[Erratum: JCAP 09, E01 (2020)]},
  \eprint{2004.01135}.

\bibitem[{\citenamefont{Flauger et~al.}(2020)\citenamefont{Flauger, Karnesis,
  Nardini, Pieroni, Ricciardone, and Torrado}}]{Flauger2020qyi}
\bibinfo{author}{\bibfnamefont{R.}~\bibnamefont{Flauger}},
  \bibinfo{author}{\bibfnamefont{N.}~\bibnamefont{Karnesis}},
  \bibinfo{author}{\bibfnamefont{G.}~\bibnamefont{Nardini}},
  \bibinfo{author}{\bibfnamefont{M.}~\bibnamefont{Pieroni}},
  \bibinfo{author}{\bibfnamefont{A.}~\bibnamefont{Ricciardone}},
  \bibnamefont{and} \bibinfo{author}{\bibfnamefont{J.}~\bibnamefont{Torrado}}
  (\bibinfo{year}{2020}), \eprint{2009.11845}.

\bibitem[{\citenamefont{Bond et~al.}(2000)\citenamefont{Bond, Jaffe, and
  Knox}}]{Bond1998qg}
\bibinfo{author}{\bibfnamefont{J.}~\bibnamefont{Bond}},
  \bibinfo{author}{\bibfnamefont{A.~H.} \bibnamefont{Jaffe}}, \bibnamefont{and}
  \bibinfo{author}{\bibfnamefont{L.}~\bibnamefont{Knox}},
  \bibinfo{journal}{Astrophys. J.} \textbf{\bibinfo{volume}{533}},
  \bibinfo{pages}{19} (\bibinfo{year}{2000}), \eprint{astro-ph/9808264}.

\bibitem[{\citenamefont{Sievers et~al.}(2003)}]{Sievers2002tq}
\bibinfo{author}{\bibfnamefont{J.}~\bibnamefont{Sievers}} \bibnamefont{et~al.},
  \bibinfo{journal}{Astrophys. J.} \textbf{\bibinfo{volume}{591}},
  \bibinfo{pages}{599} (\bibinfo{year}{2003}), \eprint{astro-ph/0205387}.

\bibitem[{\citenamefont{Verde et~al.}(2003)}]{Verde2003ey}
\bibinfo{author}{\bibfnamefont{L.}~\bibnamefont{Verde}} \bibnamefont{et~al.}
  (\bibinfo{collaboration}{WMAP}), \bibinfo{journal}{Astrophys. J. Suppl.}
  \textbf{\bibinfo{volume}{148}}, \bibinfo{pages}{195} (\bibinfo{year}{2003}),
  \eprint{astro-ph/0302218}.

\bibitem[{\citenamefont{Hamimeche and Lewis}(2008)}]{Hamimeche2008ai}
\bibinfo{author}{\bibfnamefont{S.}~\bibnamefont{Hamimeche}} \bibnamefont{and}
  \bibinfo{author}{\bibfnamefont{A.}~\bibnamefont{Lewis}},
  \bibinfo{journal}{Phys. Rev. D} \textbf{\bibinfo{volume}{77}},
  \bibinfo{pages}{103013} (\bibinfo{year}{2008}), \eprint{0801.0554}.

\bibitem[{\citenamefont{Handley
  et~al.}(2015{\natexlab{a}})\citenamefont{Handley, Hobson, and
  Lasenby}}]{Handley2015fda}
\bibinfo{author}{\bibfnamefont{W.}~\bibnamefont{Handley}},
  \bibinfo{author}{\bibfnamefont{M.}~\bibnamefont{Hobson}}, \bibnamefont{and}
  \bibinfo{author}{\bibfnamefont{A.}~\bibnamefont{Lasenby}},
  \bibinfo{journal}{Mon. Not. Roy. Astron. Soc.}
  \textbf{\bibinfo{volume}{450}}, \bibinfo{pages}{L61}
  (\bibinfo{year}{2015}{\natexlab{a}}), \eprint{1502.01856}.

\bibitem[{\citenamefont{Handley
  et~al.}(2015{\natexlab{b}})\citenamefont{Handley, Hobson, and
  Lasenby}}]{Handley2015vkr}
\bibinfo{author}{\bibfnamefont{W.}~\bibnamefont{Handley}},
  \bibinfo{author}{\bibfnamefont{M.}~\bibnamefont{Hobson}}, \bibnamefont{and}
  \bibinfo{author}{\bibfnamefont{A.}~\bibnamefont{Lasenby}},
  \bibinfo{journal}{Mon. Not. Roy. Astron. Soc.}
  \textbf{\bibinfo{volume}{453}}, \bibinfo{pages}{4385}
  (\bibinfo{year}{2015}{\natexlab{b}}), \eprint{1506.00171}.

\bibitem[{\citenamefont{Torrado and Lewis}(2020)}]{torrado2020cobaya}
\bibinfo{author}{\bibfnamefont{J.}~\bibnamefont{Torrado}} \bibnamefont{and}
  \bibinfo{author}{\bibfnamefont{A.}~\bibnamefont{Lewis}}
  (\bibinfo{year}{2020}), \eprint{2005.05290}.

\bibitem[{\citenamefont{Lewis}(2019)}]{Lewis2019xzd}
\bibinfo{author}{\bibfnamefont{A.}~\bibnamefont{Lewis}} (\bibinfo{year}{2019}),
  \eprint{1910.13970}.

\bibitem[{\citenamefont{Toubiana et~al.}(2020{\natexlab{b}})}]{Toubiana2020drf}
\bibinfo{author}{\bibfnamefont{A.}~\bibnamefont{Toubiana}} \bibnamefont{et~al.}
  (\bibinfo{year}{2020}{\natexlab{b}}), \eprint{2010.06056}.

\bibitem[{\citenamefont{Jani et~al.}(2019)\citenamefont{Jani, Shoemaker, and
  Cutler}}]{Jani2019ffg}
\bibinfo{author}{\bibfnamefont{K.}~\bibnamefont{Jani}},
  \bibinfo{author}{\bibfnamefont{D.}~\bibnamefont{Shoemaker}},
  \bibnamefont{and} \bibinfo{author}{\bibfnamefont{C.}~\bibnamefont{Cutler}},
  \bibinfo{journal}{Nature Astron.} \textbf{\bibinfo{volume}{4}},
  \bibinfo{pages}{260} (\bibinfo{year}{2019}), \eprint{1908.04985}.

\bibitem[{\citenamefont{Toubiana
  et~al.}(2020{\natexlab{c}})\citenamefont{Toubiana, Marsat, Babak, Baker, and
  Dal~Canton}}]{Toubiana2020cqv}
\bibinfo{author}{\bibfnamefont{A.}~\bibnamefont{Toubiana}},
  \bibinfo{author}{\bibfnamefont{S.}~\bibnamefont{Marsat}},
  \bibinfo{author}{\bibfnamefont{S.}~\bibnamefont{Babak}},
  \bibinfo{author}{\bibfnamefont{J.}~\bibnamefont{Baker}}, \bibnamefont{and}
  \bibinfo{author}{\bibfnamefont{T.}~\bibnamefont{Dal~Canton}},
  \bibinfo{journal}{Phys. Rev. D} \textbf{\bibinfo{volume}{102}},
  \bibinfo{pages}{124037} (\bibinfo{year}{2020}{\natexlab{c}}),
  \eprint{2007.08544}.

\bibitem[{\citenamefont{Dayal et~al.}(2019)\citenamefont{Dayal, Rossi,
  Shiralilou, Piana, Choudhury, and Volonteri}}]{101093mnrasstz897}
\bibinfo{author}{\bibfnamefont{P.}~\bibnamefont{Dayal}},
  \bibinfo{author}{\bibfnamefont{E.~M.} \bibnamefont{Rossi}},
  \bibinfo{author}{\bibfnamefont{B.}~\bibnamefont{Shiralilou}},
  \bibinfo{author}{\bibfnamefont{O.}~\bibnamefont{Piana}},
  \bibinfo{author}{\bibfnamefont{T.~R.} \bibnamefont{Choudhury}},
  \bibnamefont{and}
  \bibinfo{author}{\bibfnamefont{M.}~\bibnamefont{Volonteri}},
  \bibinfo{journal}{Monthly Notices of the Royal Astronomical Society}
  \textbf{\bibinfo{volume}{486}}, \bibinfo{pages}{2336} (\bibinfo{year}{2019}),
  ISSN \bibinfo{issn}{0035-8711},
  \eprint{https://academic.oup.com/mnras/article-pdf/486/2/2336/28493324/stz897.pdf},
  \urlprefix\url{https://doi.org/10.1093/mnras/stz897}.

\bibitem[{\citenamefont{Marcoccia et~al.}(2021)}]{marcoccia2021}
\bibinfo{author}{\bibfnamefont{P.}~\bibnamefont{Marcoccia}}
  \bibnamefont{et~al.}, \bibinfo{journal}{Manuscript in Preparation}
  (\bibinfo{year}{2021}).

\bibitem[{\citenamefont{Khan et~al.}(2016)\citenamefont{Khan, Husa, Hannam,
  Ohme, P\"urrer, Forteza, and Boh\'e}}]{phenomd}
\bibinfo{author}{\bibfnamefont{S.}~\bibnamefont{Khan}},
  \bibinfo{author}{\bibfnamefont{S.}~\bibnamefont{Husa}},
  \bibinfo{author}{\bibfnamefont{M.}~\bibnamefont{Hannam}},
  \bibinfo{author}{\bibfnamefont{F.}~\bibnamefont{Ohme}},
  \bibinfo{author}{\bibfnamefont{M.}~\bibnamefont{P\"urrer}},
  \bibinfo{author}{\bibfnamefont{X.~J.} \bibnamefont{Forteza}},
  \bibnamefont{and} \bibinfo{author}{\bibfnamefont{A.}~\bibnamefont{Boh\'e}},
  \bibinfo{journal}{Phys. Rev. D} \textbf{\bibinfo{volume}{93}},
  \bibinfo{pages}{044007} (\bibinfo{year}{2016}),
  \urlprefix\url{https://link.aps.org/doi/10.1103/PhysRevD.93.044007}.

\end{thebibliography}
\end{document}